\documentclass[useAMS,usenatbib]{mnras}
\usepackage{epsfig,rotate,graphicx}
\usepackage[fleqn]{amsmath}
\usepackage{subfig}
\usepackage{lscape}
\usepackage{bm}
\usepackage{tikz}
\usepackage{paralist}
\usetikzlibrary{shapes.geometric, arrows}

\newcommand{\p}{\partial}

\newcommand{\be}{\begin{equation}}
\newcommand{\ee}{\end{equation}}
\newcommand{\gtrsim}{\;\raisebox{-.8ex}{$\buildrel{\textstyle>}\over\sim$}\;}
\newcommand{\lesssim}{\; \raisebox{-.8ex}{$\buildrel{\textstyle<}\over\sim$}\;}

\newcommand{\avg}[1]{\left\langle #1 \right\rangle}

\newcommand{\ii}{\mathrm{i}}

\newcommand{\rhod}{\rho_\mathrm{d}}
\newcommand{\rhog}{\rho_\mathrm{g}}

\newcommand{\vd}{\bm{v}_\mathrm{d}}
\newcommand{\vg}{\bm{v}_\mathrm{g}}
\newcommand{\tstop}{t_\mathrm{s}}
\newcommand{\fdust}{f_\mathrm{d}}
\newcommand{\OmK}{\Omega_\mathrm{K}}
\newcommand{\OmF}{\Omega_\mathrm{K0}}

\newcommand{\Hgas}{H_\mathrm{g}}
\newcommand{\Hdust}{H_\mathrm{d}}

\newcommand{\Tstop}{T_\mathrm{stop}}

\newcommand{\wtil}[1]{\widetilde{#1}}
\newcommand{\metal}{\Sigma_\mathrm{d}/\Sigma_\mathrm{g}}
\newcommand{\Sigd}{\Sigma_\mathrm{d}}
\newcommand{\Sigg}{\Sigma_\mathrm{g}}
\defcitealias{lin17}{LY17}

\DeclareMathOperator{\real}{Re}
\DeclareMathOperator{\imag}{Im}

\usepackage{array,booktabs,tabularx}
\newcolumntype{R}{>{\centering\arraybackslash}X} 

\title[Dusty disc-planet simulations]{Dusty disc-planet interaction with dust-free simulations}  
\author[Chen \& Lin]{Jhih-Wei Chen\thanks{b04901165@ntu.edu.tw}, Min-Kai
  Lin\thanks{mklin@asiaa.sinica.edu.tw} 
\\Institute
  of Astronomy and Astrophysics, Academia Sinica, Taipei 10617, Taiwan   
} 

\begin{document}

\maketitle
\begin{abstract}
  Protoplanets may be born into dust-rich environments if
  planetesimals formed through streaming or gravitational 
  instabilities, or if the protoplanetary disc is undergoing 
  mass loss due to disc winds or photoevaporation. Motivated by
  this possibility, we explore the interaction between low mass
  planets and dusty protoplanetary discs with focus on disc-planet
  torques. We implement \citeauthor{lin17}'s newly developed, purely hydrodynamic model of dusty gas into the
  \textsc{Pluto} code to simulate dusty protoplanetary discs with an
  embedded planet. We find that for imperfectly coupled dust and high
  metallicity, e.g. Stokes number $10^{-3}$ and dust-to-gas ratio 
  $\Sigd/\Sigg = 0.5$, a `bubble' develops inside the
  planet's co-orbital region, which introduces unsteadiness in the
  flow. The resulting disc-planet torques sustain large amplitude oscillations { that persists} well
  beyond that in simulations with perfectly coupled dust or low
  dust-loading, where co-rotation torques are always damped. 
  We show that the desaturation of the co-rotation 
  torques by finite-sized particles is related to potential vorticity 
  generation from the misalignment of dust and gas densities. We
  briefly discuss possible implications for the orbital evolution of
  protoplanets in dust-rich discs. We also demonstrate
  \citeauthor{lin17}'s dust-free framework reproduces previous results
  { pertaining to dusty protoplanetary discs, }
  including 
dust-trapping by pressure bumps, dust settling, and the streaming instability.  
\end{abstract}


\section{Introduction}\label{intro}
%

The gravitational interaction between protoplanets and their   
protoplanetary disc, which leads to the planet's 
orbital evolution, likely plays a key role in shaping the   
architecture of planetary systems
\citep{baruteau13b,baruteau16b}. It is thus necessary to understand
how realistic disc conditions affect planet-disc interaction. 

Since the pioneering study of \cite{goldreich80} almost four decades
ago, the disc-planet problem has been 
extended to include many physical effects expected in protoplanetary
discs. These include turbulence 
\citep[e.g.][]{rnelson03,baruteau11c,uribe11,stoll17} or lack thereof
\citep[e.g.][]{rafikov02,li09,fung17}; large-scale magnetic fields
\citep[e.g.][]{terquem03,guilet13,mcnally17}; 
dynamical torques 
\citep[e.g.][]{masset03,peplinski08b,paardekooper14}; disc 
self-gravity \citep[e.g.][]{baruteau08,zhang08,baruteau11}; 
planet-induced instabilities 
\cite[e.g.][]{koller03,li05,lin11,fu14}; and non-isothermal 
discs
\citep[e.g.][]{paardekooper06,paardekooper08,baruteau08b,paardekooper10b,masset10,llambay15},  
to name a few. While these studies present important 
generalizations, they have been limited to purely gaseous discs.      

However, protoplanetary discs are dusty \citep{testi14}. It is thus
natural to consider dusty disc-planet interaction 
\citep[e.g.][]{paardekooper04,fouchet07,zhu12b,ayliffe12,dipierro16,dipierro17}.   
Recent efforts in this direction have been largely motivated
by observations of dusty sub-structures in protoplanetary discs
\citep[e.g.][]{marel13,alma15,andrews16,plas17,cieza17,boekel17}; as
disc-planet interaction is a potential mechanism to generate dust 
rings  \citep{dipierro15, picogna15,rosotti16,jin16, dong17, bae17}
and dust clumps 
(\citealt{lyra09b,zhu14,fu14b,bae16}) observed in these systems. 
These studies thus focus on the dust morphology instead of the 
planet's response to the dusty disc. This is justified if one assumes
typical dust-to-gas mass ratios of $\sim 1\%$, then disc-planet torques
are not expected to differ from pure gas models { \citep[but see][for a recent exception]{llambay18}.} 

On the other hand, planet migration in discs with high dust-to-gas
ratios has not been studied in detail. This, however, appears to be a 
natural limit to consider, since planetesimal formation in the first place may
require dust-rich environments \citep{johansen14}. Examples include
gravitational instability \citep{goldreich73} and the streaming
instability, which typically require local dust-to-gas ratios  
$> 1$ to operate efficiently \citep{youdin05a,johansen07}. Several
processes can enhance the dust-to-gas ratio, including vertical dust
settling \citep{takeuchi02}; gas mass loss 
due to photoevaporation \citep{alexander14} or magnetic winds
\citep{bai16}; and feedback of dust drag on the gas
\citep{gonzalez17,kanagawa17}.  


Motivated by these considerations, we explore the interaction between  
dust-rich discs and embedded planets with focus on disc-planet
torques. In order to connect with previous studies of planet migration  
in pure gas discs, it is desirable to model dusty gas as a single
fluid. Such an approach was recently presented by \cite{laibe14} and
\cite{price15}, and further developed by \citet[][hereafter \citetalias{lin17}]{lin17}. 

\citetalias{lin17} showed that for sufficiently small dust particles 
mixed with polytropic gas, the dusty-gas system is equivalent to   
pure-gas dynamics subject to cooling/heating. 
Since gaseous disc-planet interaction has been well-studied, 
this connection allows us 
to understand new results obtained from dusty discs in terms of
established results based on pure gas discs.  
This work presents the first { numerical} application of 
\citetalias{lin17}'s framework. 

This paper is organized as follows. In \S\ref{model} we first review  
\citetalias{lin17}'s dust-free description of dusty protoplanetary
discs and its numerical implementation. In \S\ref{2d_eqm} we describe
equilibrium 2D disc models. In 
\S\ref{dust_disc_planet}  we present numerical simulations of low-mass planets
embedded in 2D dusty discs with emphasis on disc-planet torques. We
find for sufficiently large particles and dust-loading, that
disc-planet torques remain oscillatory for extended periods of time
compared to small particles and/or dust-to-gas ratios. 
We summarise and discuss potential implications of these results in 
\S\ref{summary}. 
In the Appendices we 
present additional simulations to demonstrate  
\citetalias{lin17}'s framework reproduces several standard results
in 2D and 3D dusty discs, including dust-trapping at planet gaps, the
streaming instability, and dust settling.



 \section{Dust-free modeling of dusty protoplanetary discs}\label{model}
In this section we review the dust-free description of   
dusty protoplanetary discs developed by \citetalias{lin17}. This is
achieved by assuming small dust particles and
polytropic gas. The governing equations for dusty gas discs can then
be written in a form similar to pure gas dynamics and 
readily implemented into existing hydrodynamic codes.   

The physical system of interest is a dusty accretion disc orbiting a
central star of mass $M_*$. Disc self-gravity, magnetic fields, and
viscosity are neglected. Cylindrical $(R,\phi, z)$ and spherical 
$(r,\theta,\phi)$ co-ordinates are centred on the star. The gas phase
has density, pressure, and velocity fields $\rhog, P,
\bm{v}_\mathrm{g}$, respectively. The disc is nearly Keplerian with 
angular velocity $\Omega\simeq \Omega_\mathrm{K}(R) \equiv
\sqrt{GM_*/R^3}$ where $G$ is  the gravitational constant.  

\subsection{Simplifying assumptions}

We consider small dust particles strongly but not necessarily perfectly coupled
to the gas. We approximate the dust particles as a single,
pressureless fluid with density $\rhod $ and velocity
$\bm{v}_\mathrm{d}$ \citep{jacquet11}. The mutual dust-gas   
drag is characterized by the stopping time $\tstop$ such that 
\begin{align}  
  \rhod\left.\frac{\p \bm{v}_\mathrm{d}}{\p t}\right|_\mathrm{drag}= -
  \rhog\left.\frac{\p \bm{v}_\mathrm{g}}{\p t}\right|_\mathrm{drag}=
  - \frac{\rhog\rhod}{\rho}\frac{\left(\bm{v}_\mathrm{d} - \bm{v}_\mathrm{g}\right)}{\tstop}. 
\end{align}
is the dust-gas friction force per unit volume, 
where  
\begin{align}
  \rho \equiv \rhog + \rhod
\end{align}
is the total density.

We consider the Epstein regime with relative stopping times 
given by 
$  \tstop = \rho_\bullet s_\bullet / \rho c_s $
\citep{weidenschilling77}, where $\rho_\bullet$, $s_\bullet$ is the
internal density and radius of a grain, respectively; and $c_s$ is the
gas sound-speed (see below). In practice, we specify the  
dimensionless stopping time or Stokes number $\Tstop$
such that 
\begin{align} 
  \tstop = 
  \frac{\rho_0c_{s0}}{\rho
    c_s}\frac{\Tstop}{\Omega_\mathrm{K0}},\label{drag_law}   
\end{align}
where subscript 0 here and
below corresponds to evaluation at the reference cylindrical radius 
$R_0$ and disc midplane $z=0$. 
Note that our definition of the stopping time is related to the 
particle stopping time $\tau_\mathrm{s}= \tstop\rho/\rhog$. 

\subsubsection{Terminal velocity approximation}
We assume the particles are sufficiently small so their stopping times
satisfy 
\begin{align}
  \tstop\OmK \ll 1. 
\end{align} 
In this limit the dust velocity $\vd$ can be obtained from the 
\emph{terminal velocity 
approximation}:
\begin{align} 
\vd = \vg + \tstop \frac{\nabla P}{\rhog}\label{term_approx}
\end{align}
\citep[][\citetalias{lin17}]{youdin05a,jacquet11,price15}. Eq. \ref{term_approx} is a key
approximation in our model. It reflects the fact that  
particles tend to drift towards pressure maxima
\citep{whipple72,weidenschilling77}.  

\subsubsection{Polytropic gas}
The second simplification we make is to consider polytropic gas, where
the thermal pressure $P$ is given by 
\begin{align}
 P =
 c_s^2(R)\rho_\mathrm{g0}\left(\frac{\rhog}{\rho_\mathrm{g0}}\right)^\xi\equiv
 K(R)\rhog^\xi, \label{poly_eos}
\end{align}
where $c_s(R)$ is prescribed below and $\xi$ is the constant polytropic
index. In practice we consider $\xi\simeq 1$ or isothermal gas. 
Eq. \ref{poly_eos} is to hold at all times. For an ideal gas  
the disc temperature $T$ can be obtained from $P = \mathcal{R}\rhog T$, 
where $\mathcal{R}$ is the gas constant, so $c_s^2 \propto T$ for
$\xi=1$.

We take vertically isothermal, power-law radial temperature profiles
so that,    
\begin{align} 
  c_s^2(R) = c_{s0}^2 \left(\frac{R}{R_0}\right)^{-q}. \label{temp_profile}
\end{align}
We also define 
\begin{align}
  \Hgas(R) \equiv \frac{c_s(R)}{\OmK(R)}. 
\end{align}
This is the pressure scale-height for a vertically isothermal, pure
gas disc. The corresponding aspect-ratio is $h_\mathrm{g}\equiv\Hgas/R$.  

By adopting the terminal velocity approximation, we
eliminate the need to solve for the dust velocity $\bm{v}_\mathrm{d}$;
and by adopting a polytropic gas we eliminate the need for a gas 
energy equation. However, we show below that the mixture 
still obeys an effective energy equation, which completes the analogy
with hydrodynamics. 

\subsection{Governing equations} 
With the above approximations the dusty accretion disc can be
described (in a rotating frame with angular velocity
$\OmF\bm{\hat{z}}$)  
by: 
\begin{align}
 & \frac{D\rho}{D t} = -\rho\nabla\cdot\bm{v},\label{masseq}\\
  &\frac{D\bm{v}}{D t} = 
  -\frac{1}{\rho}\nabla P - \nabla \Phi - 2\OmF\bm{\hat{z}}\times\bm{v},\label{momeq}\\ 
  & \frac{D\fdust}{Dt} =
  -\frac{1}{\rho}\nabla\cdot\left(\fdust\tstop\nabla P\right) \label{dusteq_prim}
\end{align}
\citepalias{lin17}, where $\bm{v} = 
\left(\rhod\vd + \rhog\vg\right)/\rho$ is the mixture's centre of mass
velocity; $\Phi$ is an effective gravitational potential; and the dust fraction $\fdust =  \rhod/\rho$ is also given via 
the dust-to-gas ratio $\epsilon = \rhod/\rhog =
\fdust/\left(1-\fdust\right)$. 

Eq. \ref{masseq}---\ref{momeq} describes the conservation of total
mass and  momentum of the mixture, respectively. Eq. \ref{dusteq_prim}
results from dust mass conservation with the terminal velocity
approximation. The diffusion-like term on the
right-hand-side of Eq. \ref{dusteq_prim} accounts for the effect of
finite dust-gas coupling. If $\tstop=0$ then dust is perfectly coupled to the gas, and
$\fdust$ is advectively conserved. When $\tstop>0$, dust can slip past the
gas and drift toward pressure maxima. 

We refer the reader to \cite{laibe14} for a derivation of these
equations in the non-rotating frame and \cite{price15} for an
implementation in Smoothed Particle Hydrodynamics. Recent applications
can be found in \cite{ragusa17} and \cite{tricco17}. We will implement 
an alternate form of these equations in grid-based hydrodynamics.  

\subsection{Gravitational potential}\label{gpot}
In the rotating frame the total 
gravitational potential is      

\begin{align}
\Phi(R,\phi,z) =& \Phi_*(R,z) + \Phi_p(R,\phi, z) -
\frac{1}{2}R^2\OmF^2. 
\end{align}
Here, 
\begin{align}
  \Phi_*(R,z)  = - \frac{GM_*}{\sqrt{R^2 + z^2}}\simeq 
 - R^2\OmK^2 \left(1 - 
    \frac{z^2}{2R^2}\right).\label{thin_pot} 
\end{align}
is the stellar potential, and the second equality is its approximate
form in a thin-disc ($|z|\ll R$)

We also allow for a planet fixed on a Keplerian circular orbit at
radius $R_0$ at the disc midplane. Placing the planet at $\phi=\pi$
in the rotating frame, the planet potential is   
\begin{align}
\Phi_p(R,\phi,z) = &-\frac{GM_p}{\sqrt{R^2 + R_0^2 - 2 R R_0 \cos{(\phi-\pi)}
    + z^2 + r_s^2}} \notag\\ 
&+\frac{GM_p}{R_0^2}R\cos{(\phi-\pi)},   
\end{align}
where $r_s$ is the softening length. 
The  
second term is the indirect potential to account for the acceleration
of the star-planet system centre of mass.  

\subsection{Dust evolution as effective energy equations}  

Eq. \ref{dusteq_prim}, which governs dust evolution, may be recast
into a more familiar form by    
noting the dust fraction can be eliminated from
the problem by using the gas equation of state, Eq. \ref{poly_eos}: 
\begin{align}
  \fdust = 1 - \frac{1}{\rho}\left(\frac{P}{K}\right)^{1/\xi}. \label{fdust_replace}
\end{align}
Then Eq. \ref{dusteq_prim} can be
converted to read 
\begin{align}
  \frac{D P}{D t} = -\xi P\nabla\cdot\bm{v}
  + P\bm{v}\cdot\nabla\ln{c_s^2} + \frac{\xi
    P}{\rhog}\nabla\cdot\left(\fdust\tstop\nabla P\right), \label{dusteq}
\end{align}
with $\rhog=\rhog(P;R)$ from Eq. \ref{poly_eos} and
$\fdust=\fdust(P,\rho;R)$ from Eq. \ref{fdust_replace}.  

{ Notice Eq. \ref{dusteq} is identical in form to the pressure equation of a pure, ideal gas with adiabatic index $\gamma=\xi$, subject to heating and cooling. However, Eq. \ref{dusteq} does not physically reflect the mixture's energy, but is just a consequence of assuming a polytropic gas together with 
dust mass conservation and the terminal velocity approximation  \citepalias{lin17}. 
Nevertheless,}  we use  Eq. \ref{dusteq} instead 
Eq. \ref{dusteq_prim} to convert the dusty problem to standard
hydrodynamics. 
Eq. \ref{masseq}---\ref{momeq} and Eq. \ref{dusteq} are 3 equations
for the unknowns $(\rho,\bm{v},P)$. Although the physical system
describes dusty gas, dust does not make an explicit appearance in the 
equations: the model is pure hydrodynamical.  

{ The main advantage of evolving the pressure via Eq. \ref{dusteq} instead of using Eq. \ref{poly_eos} is practical ease in the numerical implementation of dust evolution, Eq. \ref{dusteq_prim} (see \S\ref{pluto_implement}). Furthermore, modeling the dusty problem in a  dust-free manner allows us to generalise pure gas phenomena to dusty discs and gain physical insight. 
Our dust-free model connects our new results on dusty disc-planet torques to previous results established for pure gas discs \citep[e.g.][]{paardekooper10b}. 
}

In the Appendices we demonstrate this hydrodynamic framework reproduces
several { other} expected results related to dust-gas interaction in
protoplanetary discs. These include radial dust drift and dust
trapping at planet gaps  (Appendix \ref{dust_trapping}); the
streaming instability (Appendix \ref{streaming}); and dust settling
(Appendix \ref{settling}). { The latter two cases consider 3D discs.}


\subsubsection{Effective total energy}       
If we define the effective total energy density $E$ with 
\begin{align}
  E \equiv \frac{P}{\xi-1} + \frac{1}{2}\rho|\bm{v}|^2 + \rho\Phi, 
\end{align}
then combining Eqs. \ref{momeq} and \ref{dusteq} gives
\begin{align}
  \frac{\p E}{\p t} + \nabla\cdot\left[(E+P)\bm{v}\right]  = & 
  \frac{P}{\xi-1}\bm{v}\cdot\nabla\ln{c_s^2} \notag\\
 &  + \frac{\xi P}{\rhog(\xi
    -1)}\nabla\cdot\left(\fdust\tstop\nabla P\right), \label{dust_energy_tot}
\end{align}
where we have assumed a time-independent gravitational potential, so
$\p_t\Phi=0$.   

{ It should be emphasized that $E$ does not represent the true total energy density of the dusty gas. Instead, $E$ is simply defined such that upon combining the momentum and mass conservation equations  with the polytropic equation of state (Eq. \ref{poly_eos}), we arrive at an evolutionary equation (\ref{dust_energy_tot}) that is similar in form to the total energy equation of a pure gas with source terms. 
} 
The first term on the
right-hand-side (RHS) is analogous to optically-thin cooling: it
reflects the fixed temperature profile imposed upon the system 
(Eq. \ref{temp_profile}). The second, diffusion-like term  
reflects particle drift relative to the gas. 


\subsubsection{Effective entropy}\label{eff_entropy}

We can also recast the thermal energy evolution (Eq. \ref{dusteq}) in terms of entropy evolution
by { defining} the mixture's effective (dimensionless) entropy as
$S = \ln{\left(P/\rho^\xi\right)}$. Then 
\begin{align}
  \frac{DS}{Dt} = \bm{v}\cdot\nabla \ln{c_s^2} +
  \frac{\xi}{\rhog}\nabla\cdot\left(\fdust\tstop\nabla
    P\right). \label{entropy_eq}
\end{align}
This form is appropriate for hydrodynamic codes that evolve the system
entropy, rather than energy. 


\subsubsection{Physical interpretation}\label{dust_mod_temp} 

Although the true disc temperature $T$ is prescribed via the equation of
state (Eq. \ref{temp_profile}), the dusty-gas mixture still obeys an
effective energy equation, converted from the dust evolutionary
equation. This indicates that the mixture possesses a `dynamic'
temperature $\widetilde{T}$ that depends on the dust-to-gas ratio.   

To see this, let us regard the mixture as a single ideal fluid with
$P= \mathcal{R} \rho \widetilde{T}$. 
As the pressure is actually given by $P  =
\mathcal{R}\rhog T$, we find 
\begin{align}
  \widetilde{T}  = \frac{T}{1+\epsilon}, \label{dynT}
\end{align} 
and recall $\epsilon = \rhod/\rhog$. 
Thus increasing the dust-to-gas ratio lowers the mixture's dynamic
temperature. This occurs because dust-loading  increases the mixture's
inertia, but it does not contribute to thermal pressure.  


The above discussion implies the effective sound speed ($\simeq
\sqrt{P/\rho}$) of a dusty disc is reduced by a factor of $\sqrt{1 +
  \epsilon}$. Thus the 
effective scale-height $\widetilde{H}$ is also reduced:  
\begin{align}
  \widetilde{H} = \frac{\Hgas}{\sqrt{1+\epsilon}},
\end{align}
and we can define the effective aspect-ratio $\widetilde{h} \equiv
\widetilde{H}/R$. The disc becomes thinner with increasing
dust-load at fixed (true) temperature. 


In \citetalias{lin17}'s model, dust-loading is interpreted as reducing the dynamic
temperature or sound-speed, so a positive particle flux is analogous
to a negative heat flux. A particle drift towards a pressure maximum
constitutes to cooling the location of the maximum. This explains why 
dust-gas drag appears in an effective energy equation 
instead of the momentum equation. 

\subsection{Potential vorticity generation}\label{pv_gen}

It is important to notice that the dusty gas mixture is
\emph{baroclinic}, because the pressure arises solely
from the gas, while the total density involves both gas and dust. This
means the potential vorticity (PV) or vortensity is generally not
conserved, because dust and gas need not be aligned if they can drift
relative to each other. 

The PV in the inertial frame is 
\begin{align}
\bm{\zeta} \equiv \frac{\nabla\times\bm{v} +
  2\Omega_\mathrm{K0}\hat{\bm{z}}}{\rho}. 
\end{align} 
It obeys \begin{align}
\frac{D\bm{\zeta}}{Dt} = \bm{\zeta}\cdot\nabla\bm{v} +
\frac{1}{\rho^3}\nabla\rho\times\nabla P.
\end{align}
The second, baroclinic source term, is generally non-zero. 
{ For our 
  equation of state $P\propto c_s^2(R)\rhog^\xi$
  (Eq. \ref{poly_eos}) we find 
\begin{align}
\nabla \rho \times \nabla P  = P \nabla \rho \times \nabla \ln c_s^2 + \frac{\xi
    P}{\rhog}\nabla\rhod \times\nabla\rhog. \label{baroclinic_expand}
\end{align}
The first term reflects baroclinity due misalignment between the (imposed)
temperature profile and the total density. In the 2D, razor-thin discs
models we consider later,  this misaligntment requires $\p_Rc_s^2\neq 0$ 
with a non-axisymmetric total density distribution,  $\p_\phi \rho\neq~0$. 
However, this source of baroclinity is negligible in our disc-planet
models. 
} 

{ Baroclinity may also arise from a misalignment between gas and
  dust densities, $\nabla \rhod \times \nabla
\rhog\neq0$, which requires a non-uniform dust-to-gas ratio. 
For our disc models below this non-uniformity --- and hence baroclinity
---   
largely develops from finite dust-gas drag (Eq. \ref{dusteq_prim}).  
}

The baroclinic nature of dusty gas was already pointed out by
\cite{laibe14}. In this work, we present a realization of this
effect in disc-planet interaction. Namely, we will show PV can be 
generation from dust-gas misalignment near the planet { due to dust-gas drift.}

\subsection{Numerical implementation}\label{pluto_implement}


Fortunately, many publicly available hydrodynamic codes solve 
the energy equation in the form of Eq. \ref{dust_energy_tot}. In this
work we use \textsc{Pluto}\footnote{\url{http://plutocode.ph.unito.it}}  
\citep{mignone07,mignone12} --- a general purpose, finite-volume
Godunov code for simulating astrophysical fluids. It can treat a
variety of non-ideal hydrodynamic effects, including energy source
terms similar to the RHS of Eq. \ref{dust_energy_tot}. Thus 
little effort is needed to convert \textsc{Pluto} into a dusty
hydrodynamics code.     

The true disc temperature profile is imposed by the 
$\nabla c_s^2$ term on the RHS of Eq. \ref{dust_energy_tot}. For this 
we enable \textsc{Pluto}'s optically-thin cooling module to update the
pressure by solving   
\begin{align}
  \frac{\p P}{\p t} = P v_R\p_R\ln{c_s^2},  \quad \Rightarrow  \quad
  \Delta P = P \left[\exp{\left(- q\frac{v_R}{R}\Delta
        t\right)}-1\right], 
\end{align}
where $\Delta P$ and $\Delta t$ is the pressure update and time-step,
respectively; and we have used Eq. \ref{temp_profile}.  

Dust-gas drag is represented by second, diffusive term on the RHS of 
Eq. \ref{dust_energy_tot}. For this we enable \textsc{Pluto}'s  
thermal conduction module. The original module updates the total energy $E$ by solving 
\begin{align}
  \frac{\p E}{\p t} = \nabla\cdot\left(\kappa_T\nabla T\right), \label{pluto_therm_cond}
\end{align}
where $T\propto P/\rho$ is the temperature field and $\kappa_T$ is the thermal 
conduction coefficient. We adapt the module by replacing the   
temperature gradient in Eq. \ref{pluto_therm_cond} with the pressure gradient,   
\begin{align}
  \nabla T \to \nabla P,
\end{align}
and setting the thermal conduction coefficient to 
\begin{align}
  \kappa_T \to \fdust \tstop, 
\end{align}
with $\fdust = \fdust(\rho, P; R)$ via Eq. \ref{fdust_replace}.
Finally, we scale the thermal conduction term by $\xi P/ \rhog (\xi-1)$, see
Eq. \ref{dust_energy_tot}. In practice this scaling factor is almost a
constant because we choose $\xi\simeq 1$. 
We use \textsc{Pluto}'s Super-Time-Stepping algorithm to
integrate the dust diffusion term.



\section{Razor-thin discs}\label{2d_eqm}


In this paper we mostly consider 2D, razor-thin discs. To obtain { the} 2D
disc equations from the full 3D models above, we replace $\rho$ by the
surface density $\Sigma$ (similarly for the dust and gas 
components) and re-interpret $P$ as the vertically-integrated
pressure. Thus, for example, the equation of state becomes
$P=K\Sigma_\mathrm{g}^\xi$. We set $v_z=0$ and the 
potential functions in \S\ref{gpot} are evaluated at
$z=0$. Alternatively, we can consider these 2D models to represent
a layer of disc material about the disc midplane. 

\subsection{Equilibrium structure}\label{eqm_struct}
We initialize the 2D disc in an axisymmetric, steady state with
total surface density $\Sigma = \Sigma(R) = \Sigma_\mathrm{g}[1 +
\epsilon(R)]$ and velocity field $v_\phi(R) = R\left[\Omega(R)-\Omega_\mathrm{K0}
\right]$, $v_R=0$. These satisfy the
equilibrium radial momentum and energy equations (without a planet)
\begin{align}
  R\Omega^2 &= R\OmK^2(R) + \frac{1}{\Sigma}\frac{d P}{d 
    R},\label{rad_eqm}\\ 
  0& = \frac{d}{d R}\left(R \fdust\tstop \frac{d P}{d 
      R}\right) \label{therm_eqm}. 
\end{align}
We set the initial gas surface density
to \begin{align} 
  \Sigma_\mathrm{g}(R)  =
  \Sigma_\mathrm{g0}\left(\frac{R}{R_0}\right)^{-p}. \label{surfden}
\end{align}
For a non-self-gravitating problem, the surface density scale
$\Sigma_\mathrm{g0}$ is arbitrary.

To obtain $\Sigma$ from $\Sigma_\mathrm{g}$ we choose
$\epsilon(R)$ to satisfy `thermal' equilibrium,
Eq. \ref{therm_eqm}. This equation implies  
\begin{align}
\frac{\epsilon}{\left(1+\epsilon\right)^2}c_s(R)\Sigma_\mathrm{g}^{\xi-1}(R)\times\frac{d\ln
P}{d\ln R}=\mathrm{constant},   
\end{align} 
where we have used the 2D version of the Epstein drag law,
Eq. \ref{drag_law}.  
For power-law discs with generally non-zero pressure gradients, 
we require the first factor to be constant. We thus obtain
$\epsilon(R)$ from  
\begin{align*}
\frac{\epsilon(R)}{\left[1+\epsilon(R)\right]^2} &=
\frac{\epsilon_0}{\left(1+\epsilon_0\right)^2} 
\left[\frac{c_{s0}}{c_s(R)}\right]\left[\frac{\Sigma_\mathrm{g0}}{\Sigma_\mathrm{g}(R)}\right]^{\xi-1}
\equiv F(R). 
\end{align*}
Notice 
\begin{align}
\frac{\left(1-\epsilon\right)}{(1+ \epsilon)^3}\frac{d\epsilon}{dR} =
\frac{dF}{dR}. \label{dFdR}
\end{align}
For typical disc models $dF/dR>0$. Then
dust-poor discs ($\epsilon <1$) have dust-to-gas ratios
increasing with $R$; whereas dust-rich discs ($\epsilon >
1$) have dust-to-gas ratios decreasing with $R$. 
Evaluating Eq. \ref{dFdR} explicitly gives 
\begin{align}
\frac{d\ln{\left(1+\epsilon\right)}}{d\ln{R}} = \frac{\epsilon}{1 -
  \epsilon}\left[ \frac{q}{2}  
+ (\xi-1)p \right].\label{dlogepsilon}
\end{align} 
The special case of constant dust-to-gas ratio requires 
\begin{align}
  q = -2(\xi -1 )p \quad \text{for
    constant $\frac{\Sigma_\mathrm{d}}{\Sigma_\mathrm{g}}$}.  
\end{align}
For nearly isothermal gas and $p$ of order unity, the disc 
temperature is close to being globally isothermal.  

Having specified $\Sigma$ we can use Eq. \ref{rad_eqm} to set the
rotation profile 
\begin{align}
\Omega(R)= \OmK\left[1 - 
    \frac{P}{\Sigma\left(R\OmK\right)^2}\left(q + \xi p\right)\right]^{1/2} \label{eqm_rotation}, 
\end{align}
where we have used Eq. \ref{temp_profile}.  

\subsection{Effective disc structure}\label{eff_struct}



Using the equilibrium conditions in \S\ref{eqm_struct}, we find 
the radial gradient of the total surface density becomes  
\begin{align}
\widetilde{p} \equiv - \frac{d\ln{\Sigma}}{d\ln{R}} =  p -
 \frac{\epsilon}{1 - 
   \epsilon}\left[\frac{q}{2} + (\xi-1)p\right]. 
\end{align}
Similarly, the radial gradient in the dynamic temperature is  
\begin{align}
\widetilde{q} \equiv -\frac{\p\ln{\widetilde{T}}}{\p\ln{R}} = q + 
\frac{\epsilon}{1 - \epsilon}\left[\frac{q}{2} 
+ (\xi-1)p\right].    
\end{align}
For constant $\epsilon$ the term in square
brackets is zero, so $(p,q)=(\widetilde{p},\widetilde{q})$ in this
special case. 



\section{Dusty disc-planet interaction}\label{dust_disc_planet}

We simulate the response of 2D, razor-thin dusty discs described above
to an embedded low-mass planet and focus on the associated disc-planet
torques. We set $\xi=1.001$ to model isothermal gas. We adopt 
$p=q=0$ as our standard disc profile, which has zero radial pressure
gradient and thus no radial drift of particles initially. We fix the dynamic 
disc temperature such that $\widetilde{h}_0=0.05$ for all
simulations by letting the gas aspect-ratio vary as 
$h_\mathrm{g0}~=~\sqrt{1+Z}\,\widetilde{h}_0$, where
$Z\equiv\metal$ is the metallicity. This counter-acts the reduction in the
effective scale-height $\widetilde{H}$ due to dust-loading. 

We consider a planet mass of $M_p =6\times 10^{-6}M_*$, which
corresponds to a $2M_{\earth}$ planet orbiting a solar mass star. The
planet is held on a fixed Keplerian circular orbit at $R_0$. The
planet potential is switched on smoothly over one orbit. We adopt a
fiducial softening length of $r_s =0.6\widetilde{H}_0\simeq 2.38r_h$,
where $r_h = (M_p/3M_*)^{1/3}R_0$ is the planet's Hill radius.     

Our simulations summarized in Table \ref{discp_runs}. We primarily
vary the Stokes number $\Tstop\equiv t_\mathrm{s0}\Omega_\mathrm{K0}$ and
metallicity $Z$ to explore the effect of imperfect dust-gas
coupling and dust-loading on disc-planet torques, respectively. We
also vary the power-law indices $p,q$ to examine the effect of the
background potential vorticity and dust-to-gas ratio profiles, and
consider the effect numerical resolution. 

\begin{table*}
  \caption{Summary of disc parameters used in disc-planet
    simulations. Comments are made relative to the fiducial run.} 
  \label{discp_runs}
  \begin{tabular}{lrrrrl}
    \hline    
    Run &  $\Sigma_\mathrm{d}/\Sigma_\mathrm{g}$ &
    $T_\mathrm{stop}$ & $(p,q)$ & $N_R\times N_\phi$ & Comment\\ 
    \hline
    Fiducial   & $0.5$  & $0      $ & $(0,0)$ & $720\times2672$ & 
    Reference simulation with perfectly-coupled particles. Fig.\ref{torques_ts0}.\\
    Decouple  & $0.5$  & $10^{-3}$ & $(0,0)$   & $720\times2672$ & 
    Partially coupled particles. Fig.\ref{torques_extend2}.\\ 
    LowDust   & $0.01$ & $10^{-3}$ & $(0,0)$   & $720\times2672$ & 
    Reduced dust-loading.  Fig.\ref{torques_vary_d2g}.\\  
    HighDust  & $1.0$  & $10^{-3}$ & $(0,0)$  & $720\times2672$ & 
    Increased dust-loading. Fig.\ref{torques_vary_d2g}.\\ 
    FlatPV    & $0.5$  & $10^{-3}$ & $(1.5,0)$ & $720\times2672$ &  Zero
    vortensity gradient (uniform dust-to-gas
    ratio). Fig.\ref{torques_flat_pv}.\\
    PPD0 & $0.5$  &  $0$ & $(1,0.5)$ &  $720\times2672$
    & Protoplanetary disc with perfectly coupled dust. Fig. \ref{torques_typical}.\\
    PPD  &  $0.5$   & $10^{-3}$ & $(1,0.5)$ &  $720\times2672$
    & Protoplanetary disc with imperfectly coupled dust. Fig. \ref{torques_typical}.\\
    LowRes    & $0.5$  & $10^{-3}$ & $(0,0)$  & $360\times1336$ &  Lowered
    numerical resolution. Fig.\ref{torques_vary_res}.\\
    HighRes   & $0.5$  & $10^{-3}$ & $(0,0)$   & $1440\times5344$ & 
    Increased numerical resolution. Fig.\ref{torques_vary_res}.\\
    \hline
  \end{tabular}
\end{table*}

\subsection{Numerical setup}



The disc domain is $R\in[0.6,1.4]R_0$ and $\phi\in[0,2\pi]$. 
We damp all variables towards their initial values in
$R\in[0.6,0.68]R_0$ and  $R\in[1.32,1.4]R_0$ on a timescale of the
local orbital period. The radial boundary ghost cells are kept at
their initial states, while azimuthal boundaries are periodic. This
setup is similar to \cite{kley12b}. 

We use $N_R$ cells logarithmically spaced in $R$ and $N_\phi$ cells
uniformly spaced in $\phi$. Our standard numerical resolution 
is such that $\widetilde{H}$ is  
resolved by $(40,20)$ cells in the radial and 
azimuthal directions, respectively.   
This is similar to the resolution adopted by
\cite{paardekooper10b}. 

We adopt standard configurations for the \textsc{Pluto} code with
linear reconstruction, the HLLC Riemann solver and second order
Runge-Kutta time integration, and enable the FARGO algorithm
\citep{masset00a,masset00b} to accelerate the simulations.

\subsection{Torque analyses}\label{torque_analysis}

The disc-on-planet torque per unit area, or torque density, is
defined as 
\begin{align} 
  \Gamma_\Sigma \equiv \Sigma\frac{\p\Phi_p}{\p\phi}, 
\end{align} 
so the total torque acting on the planet is 
\begin{align}
\Gamma = \iint {\Gamma}_\Sigma R dR d\phi. 
\end{align} 
This integration includes the entire disc, i.e. no cut-off is applied 
near the planet. 
We also define the torque per unit radius as 
\begin{align}
\frac{\p\Gamma}{\p R} = R \int \Gamma_\Sigma d\phi. 
\end{align}

We find the effect of dust  is to introduce
unsteadiness in torques originating from the planet's co-orbital region. 
To focus the torque analysis 
in this region, we define the time-averaged torque per unit radius 
\begin{align}
\avg{\frac{\p\Gamma}{\p R}} =
\frac{1}{T_\mathrm{avg}}\int_{T_\mathrm{avg}} \frac{\p \Gamma}{\p R} dt. 
\label{background_torque}
\end{align}
The time interval $T_\mathrm{avg}$ is the oscillation period for the
total torque (see \S\ref{fid_run}). This time-averaged torque
essentially captures the steady Lindblad or wave torques outside the co-orbital region; removing it  
thus allow us to focus on the co-rotation torques.

We compare the measured torques to a semi-analytic torque formula
adapted from that based on pure gas simulations developed by 
\cite{paardekooper10b}. In the dust-free limit our models correspond  
to locally isothermal discs, we thus modify relevant 
torque formula from \citeauthor{paardekooper10b} to: 
\begin{align}
\frac{\Gamma}{\Gamma_\mathrm{ref}}= &\hat{\Gamma}_\mathrm{L}+ 
\hat{\Gamma}_\mathrm{c}+ \hat{\Gamma}_\mathrm{hs}  \notag\\
=& -\left(2.5 - 0.5\wtil{q} -
  0.1\wtil{p}\right)\left(\frac{0.4}{b}\right)^{0.71} \notag\\ &-
  1.4\wtil{q}\left(\frac{0.4}{b}\right)^{1.26} + 1.1 
    \left(\frac{3}{2} - \wtil{p}\right)
    \left(\frac{0.4}{b}\right)\label{loc_iso_torque} 
\end{align}
where 
$b \equiv r_s/\widetilde{H}_0$,  
and the torque normalization is 
\begin{align}
\Gamma_\mathrm{ref}=
\Sigma_0R_0^4\Omega_\mathrm{K0}^2\left(\frac{M_p}{M_*}\right)^2\left(
  \frac{1}{\widetilde{h}_0^2}
\right).    
\end{align} 
The expressions above are identical in form to
\citeauthor{paardekooper10b}'s Eq. 49, but { accounts}
for dust-loading in $\Sigma_0$, $\widetilde{H}_0$, 
$\widetilde{h}_0$ (\S\ref{dust_mod_temp}), and { the} dust-modified 
effective disc structure (\S\ref{eff_struct}); although there is some
uncertainty as to whether the true or dynamic temperature profile
should be used (S.-J. Paardekooper, private communication). 
Eq. \ref{loc_iso_torque} includes the
(normalized) Lindblad or wave torque $\Gamma_\mathrm{L}$; the linear
co-rotation torque $\Gamma_\mathrm{c}$; and the non-linear horseshoe drag
$\Gamma_\mathrm{hs}$. Note that $\Gamma_\mathrm{c}=0$ in our standard
disc model with $p=q=0$ and constant dust-to-gas ratio.


%

\subsection{Perfectly coupled dust}\label{fid_run}


We begin with a fiducial run with perfectly coupled dust 
($\Tstop=0$) but high dust-loading, $Z=0.5$. 
In this case the system is truly a single fluid  and the
only effect of dust is to lower the dynamic temperature of the mixture
when compared to the pure gas limit. 

\begin{figure}
\includegraphics[width=\linewidth,clip=true,trim=0cm 0cm 0cm
  0cm]{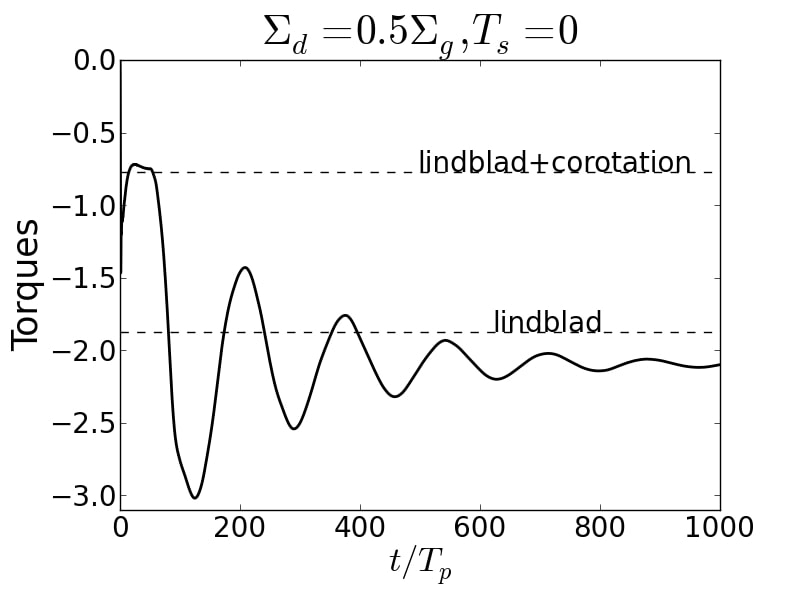}
  \caption{Disc-on-planet torques, normalized by
    $\xi_\mathrm{ref}$, for a $2M_{\earth}$ planet in the
    fiducial run  
  with dust-to-gas ratio $Z=0.5$ and stopping time
  $T_\mathrm{stop}=0$ (solid lines).   
  Horizontal lines are semi-analytic torque values adapted from \protect\cite{paardekooper10b}, where the upper (lower)
  line corresponds to the total (Lindblad) torques. 
  }\label{torques_ts0}
\end{figure}

Fig. \ref{torques_ts0} show that resulting total torque agrees well with
Eq. \ref{loc_iso_torque} for $\lesssim 50P_0$ when the horseshoe drag
remains unsaturated \citep{paardekooper10b}; and for $t\to\infty$ when
the horseshoe drag saturates due to mixing of the co-orbital region,
leaving only the Lindblad torques.    

This experiment demonstrates that for sufficiently small particles,
disc-planet torques are unaffected once dust-loading is accounted for
in the torque normalizations, namely in increasing the total  
surface density and in reducing the disc aspect-ratio. 

\subsection{Partially coupled dust and torque oscillations}  

\begin{figure}
\includegraphics[width=\linewidth,clip=true,trim=0cm 0cm 0cm
  0cm]{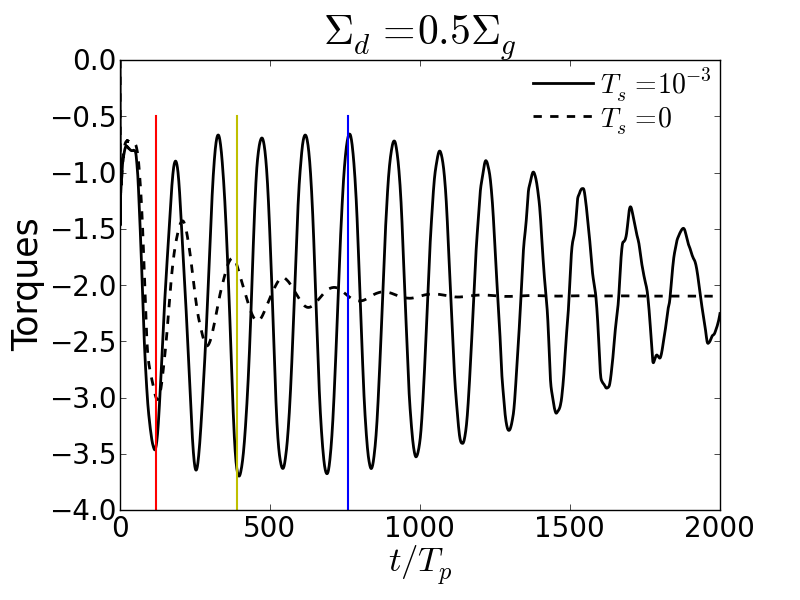}
  \caption{Disc-on-planet torques for a $2M_{\earth}$ planet in a disc
  with dust-to-gas ratio $Z=0.5$ and stopping time
  $T_\mathrm{stop}=0$ (dashed) and
  $T_\mathrm{stop}=10^{-3}$ (solid).  
  The three vertical lines correspond to representative timestamps 
  in Fig. \protect\ref{torques_radial_cmp}---\protect\ref{una_box_series}. 
  \label{torques_extend2}}
\end{figure}

We now allow the dust to be slightly decoupled from the gas by setting
$\Tstop=10^{-3}$. This decoupled run is compared to the fiducial case
in Fig. \ref{torques_extend2}. We also extended the
simulations to $t=2000P_0$ to compare their long term evolution.  

\par
In both cases the average torque is close to the Lindblad-only values. 
For $\Tstop=0$ the torque oscillations immediately damp towards the Lindblad-only values, and reaches a steady state by the end of the simulation. However, in the decoupled run with $T_\mathrm{stop}=10^{-3}$, the oscillations are sustained 
up to $\sim 1000$ orbits before starting to damp. In fact, 
the oscillation amplitudes reach the unsaturated (full) torque 
values. The torque remains oscillatory throughout the simulation, but will likely eventually converge to that of the fiducial, perfectly coupled run.   

\begin{figure}
\includegraphics[width=\linewidth,clip=true,trim=0cm 0cm 0cm
  0cm]{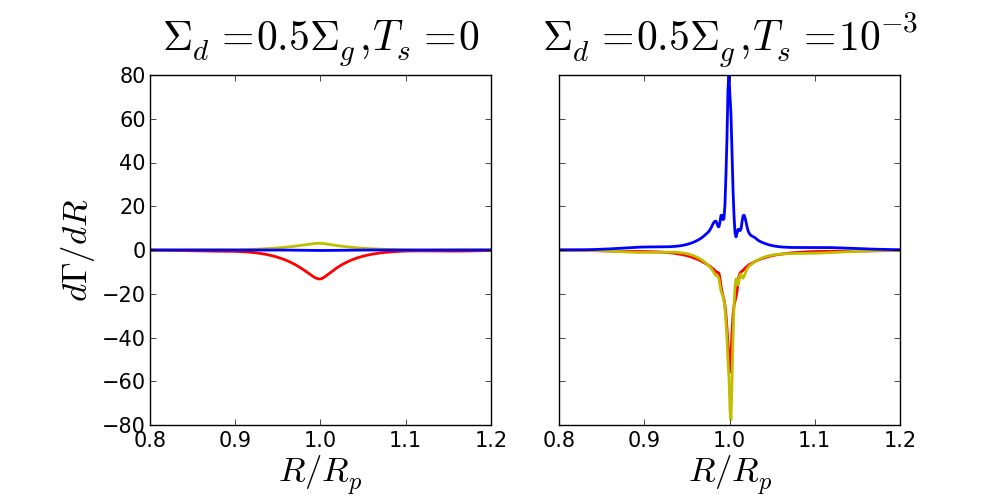}
  \caption{Disc-on-planet torques per unit radius, subtracted by
    the Lindblad torques, for a $2M_{\earth}$ planet in a disc with
    dust-to-gas ratio $Z=0.5$ and stopping time $T_\mathrm{stop}=0$
    (left) and $T_\mathrm{stop}=10^{-3}$ (right).
    The three timestamps refer to Fig. \ref{torques_extend2}.
    \label{torques_radial_cmp}}
\end{figure}

Fig. \ref{torques_radial_cmp} compares the torque distributions  
at the representative times shown in Fig. \ref{torques_extend2}. 
Following the procedure outline in \S\ref{torque_analysis}, we subtract the
steady Lindblad torques to highlight the difference in the co-rotation region. The figure shows that even a slight dust-gas decoupling can generate and sustain co-rotation torques, leading to oscillatory behaviour.


\begin{figure}
\includegraphics[width=\linewidth,clip=true,trim=0cm 0cm 0cm
  0cm]{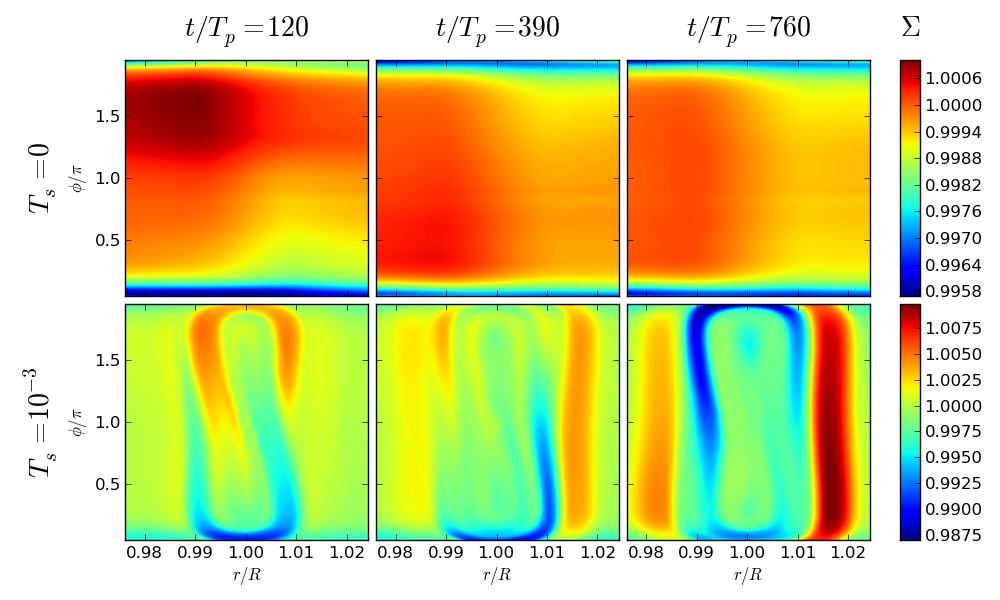}
  \caption{Normalized surface density for a $2M_{\earth}$ planet in a disc
  with dust-to-gas ratio $\epsilon_0=0.5$ and stopping time
  $T_\mathrm{stop}=0$ (upper) and $T_\mathrm{stop}=10^{-3}$ (lower).
  The three timestamps refer to Fig.  \ref{torques_extend2}. The 
  planet is located at the vertical boundaries and the co-orbital flow
  direction is clockwise. Thus the flow near the bottom (top) is ahead
  (behind) the planet.  
  \label{rho_box_series}}
\end{figure}

In Fig. \ref{rho_box_series} we compare the surface density evolution  
in the co-orbital region. Although the surface density perturbations
are small, as expected for a low mass planet, they are nevertheless
distinct depending on $\Tstop$. The $\Tstop=0$ fiducial run has a
relatively smooth surface density distribution. However, in 
the $\Tstop=10^{-3}$ decoupled run, we find a 
`bubble' of under-density (blue in the figure) develops just inside the boundary of the
co-orbital region. As it librates, it  introduces periodic front-back
surface density asymmetries about the planet, and hence sustains a  
co-rotation torque. Furthermore, we find over-dense rings develop outside the
co-orbital region, but not in the perfectly-coupled case.  

 \subsubsection{Potential vorticity generation from
   dust-gas misalignment} 
\begin{figure}
\includegraphics[width=\linewidth,clip=true,trim=0cm 0cm 0cm
  0cm]{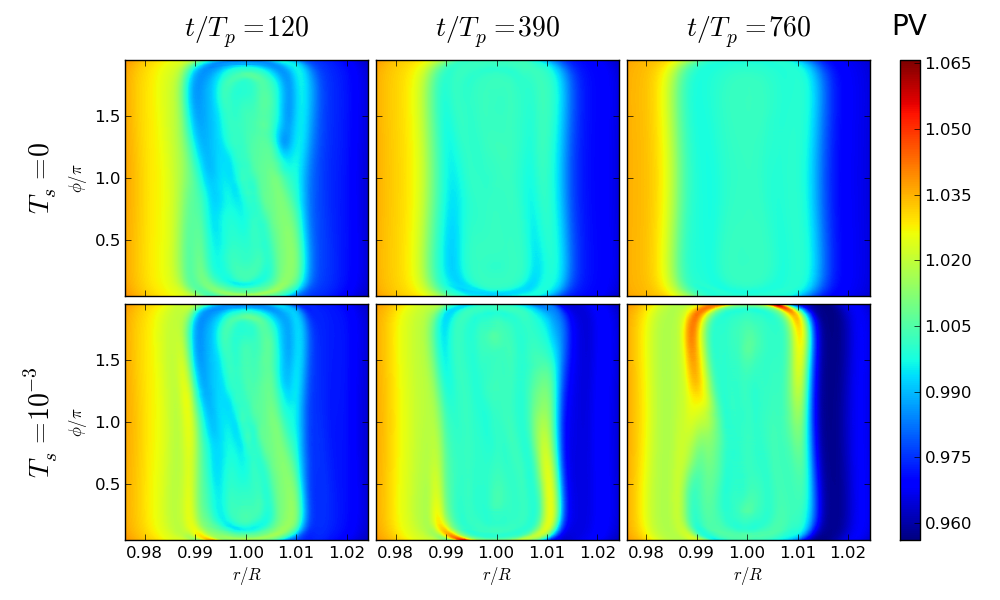}
  \caption{Potential vorticity distribution (PV) for a $2M_{\earth}$ 
  planet in a disc with dust-to-gas ratio $Z=0.5$ and stopping time
  $T_\mathrm{stop}=0$ (upper) and $T_\mathrm{stop}=10^{-3}$ (lower).
  The three timestamps refer to Fig. \protect\ref{torques_extend2}.
  \label{pv_box_series}} 
\end{figure}

We note that in pure gas discs 
the torques originating from the co-orbital region, specifically the horseshoe drag,    
is know to be related to the potential vorticity (PV) gradient 
across the co-rotation region \citep[][ see also Eq. \ref{loc_iso_torque}]{paardekooper10b}. 
We thus compare
the PV  distribution between the fiducial and decoupled runs in
Fig. \ref{pv_box_series}. Here the PV is defined as $\zeta =
\left(\hat{\bm{z}}\cdot\nabla\times\bm{v} + 
2\Omega_\mathrm{K0}\right)/\Sigma$.   

\par
  Generally, the PV in the co-orbital region undergoes phase 
  mixing, which leads to the saturation or vanishing of the
  co-rotational torque \citep{balmforth01a}. This is the case for
  $\Tstop=0$. However, for $\Tstop=10^{-3}$ we find a PV `blob' 
  develops at the boundary of the co-orbital region, though PV still 
  mixes in the interior. The PV blob corresponds to the 
  surface density bubble identified 
  above. 

  Interestingly, we find the blob
  experiences an increase in PV whenever it undergoes horseshoe turns
  as it encounters the planet. This means that when it completes a
  libration period, it experiences two boosts in PV, but it returns to
  its original orbital radius. We expect no net change in the blob's vorticity, which should be close to the local Keplerian value. 
  Then since $\zeta\propto \Sigma^{-1}$
  has increased, its surface density must drop, creating a bubble.



\begin{figure}
\includegraphics[width=\linewidth,clip=true,trim=0cm 1.3cm 0cm
  0cm]{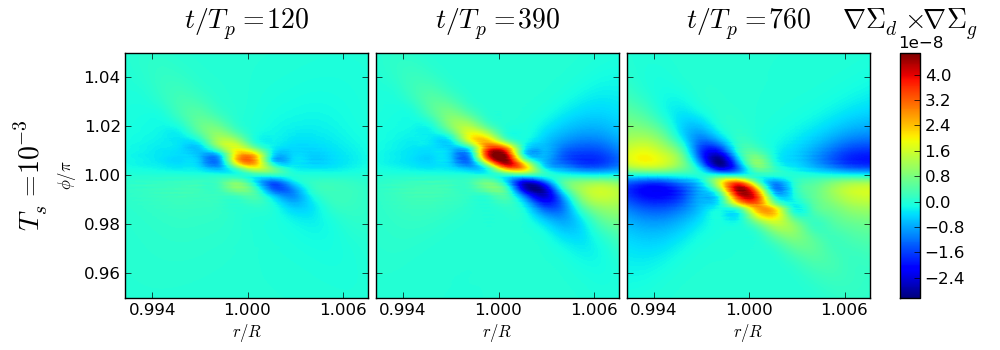}\\
\includegraphics[width=\linewidth,clip=true,trim=0cm 0cm 0cm
  0cm]{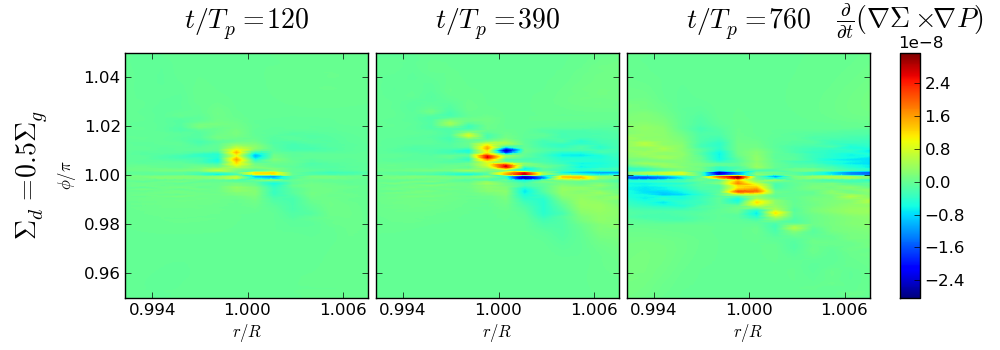}
  \caption{Baroclinic vorticity source 
    $\hat{\bm{z}}\cdot\left(\nabla\Sigd\times\nabla\Sigg\right)$ and its time derivative in the
    decoupled run with 
    $Z=0.5$ and stopping time $T_\mathrm{stop}=10^{-3}$. Here, the planet is
    placed at the centre in each plot.  
    The three timestamps refer to Fig. \ref{torques_extend2}. 
    \label{una_box_series}}
\end{figure}

In a 2D disc without shocks, the only source of PV comes from 
pressure-density misalignment, which is proportional to  
$\nabla\Sigd\times\nabla\Sigg$ (\S\ref{pv_gen}). We plot this PV 
source in upper panels of Fig. \ref{una_box_series} for the decoupled run. By
comparing the snapshots with Fig. \ref{pv_box_series}, we see the
PV source is positive when the PV blob encounters the planet 
on a horseshoe turn, so PV is indeed generated through dust-gas 
misalignment near the planet.  

Note that $\nabla\Sigd\times\nabla\Sigg = \bm{0}$ in the initial disc.
However, once the system evolves, this PV source may become non-zero
         as dust and gas drift relative to each other. We expect this
         effect to be significant near the planet as a pressure bump
         can be created by the planet's potential well. Using the 2D
         versions of Eq. \ref{masseq} and \ref{dusteq}, with $\xi=1$
         and constant $c_s$ we find     
\begin{align}
  \frac{\p}{\p t}\left(\nabla\Sigma\times\nabla P\right)
  =& \left(\frac{\p}{\p t}\nabla\Sigma\right)\times\nabla P  +
  \nabla\Sigma\times\frac{\p}{\p t}\nabla P \notag\\
  =& \nabla P \times \nabla\left[\nabla\cdot\left(\Sigma\bm{v}\right)\right]
  +  \nabla\left[\nabla\cdot\left(P\bm{v}\right)\right]\times\nabla\Sigma \notag\\
  &+c_s^2\nabla\Sigma\times\nabla\left[\nabla\cdot\left(\fdust\tstop\nabla
    P\right)\right].
    \label{baroclinic_dt}
\end{align}
We plot the right hand side of Eq. \ref{baroclinic_dt} in the lower
    panels of Fig. \ref{una_box_series}. (Interestingly, we find the
    last term is sub-dominant.) It is indeed positive
    (negative) when the baroclinic source is positive (negative), for
    a PV surplus (deficit). This suggests a positive feedback loop in
    generating the PV blob. The blob does not grow indefinitely,
    however, probably because of numerical diffusion (see
    \S\ref{resolution}).  



%

\subsubsection{Final disc structure}\label{final_structure}

\begin{figure}
\includegraphics[width=\linewidth,clip=true,trim=0cm 0cm 0cm
  0cm]{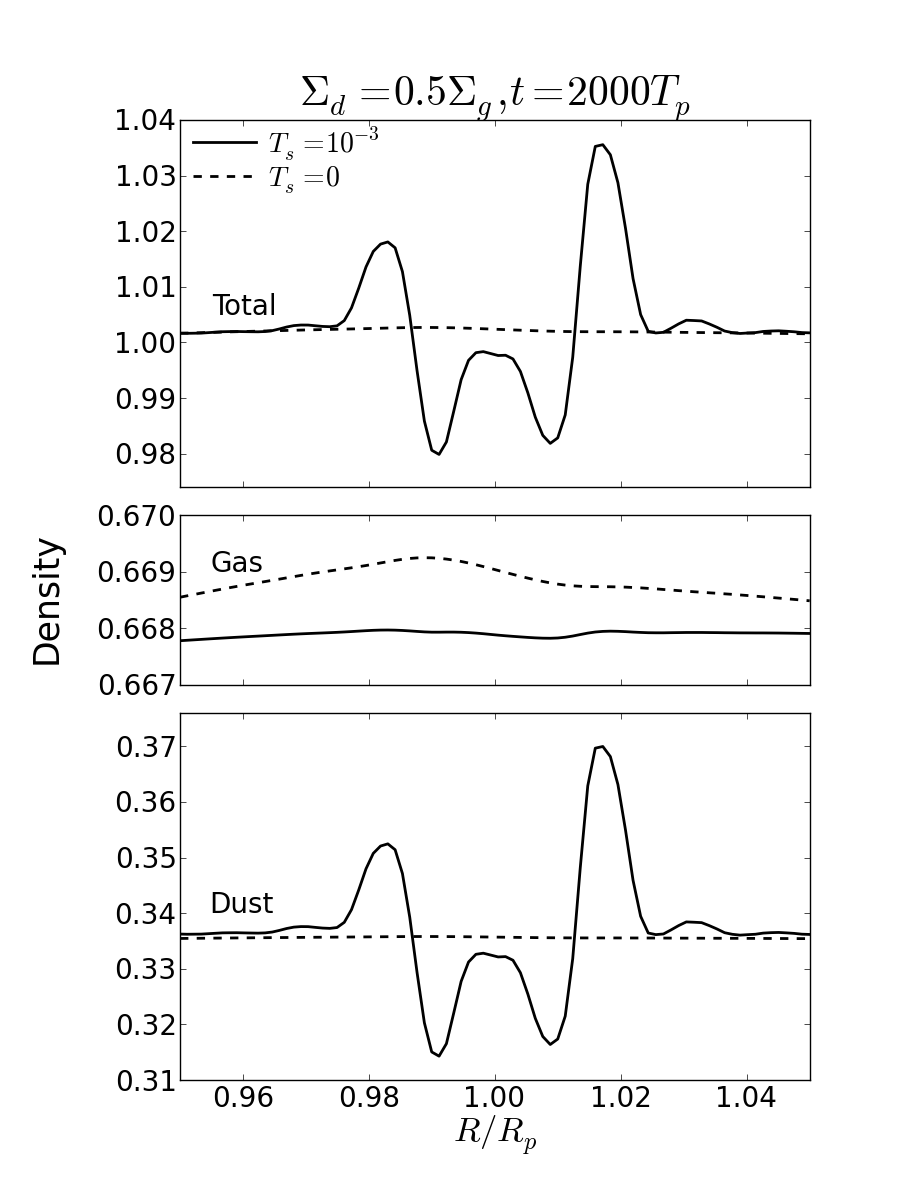}
  \caption{
  Normalized total (top), gas (middle), and dust (bottom) surface
  density distributions for a $2M_{\earth}$ 
  planet in a disc with initial dust-to-gas ratio $Z=0.5$ and stopping time 
  $T_\mathrm{stop}=10^{-3}$ (solid) and $T_\mathrm{stop}=0$ (dashed).
  \label{radial_rho_vary_ts}}
\end{figure}

In Fig. \ref{radial_rho_vary_ts} we compare the gas and dust surface density distributions
at $t=2000P_0$. For $\Tstop=0$ the dust-to-gas ratio is constant ($Z=0.5$) so the
dust distribution is simply a rescaling of the gas profile. For $\Tstop=10^{-3}$, however, a double gap 
develops in the dust. Although the dust gap is shallow, it is still significant compared to the gas,
which only show small fluctuations. This result is consistent with recent studies  
show that low mass planets can open dust gaps while leaving the gas relatively unperturbed \citep{dipierro16,dipierro17}. 

\subsection{Dust-loading and vortex instability}\label{dust_load}

\begin{figure}
\includegraphics[width=\linewidth,clip=true,trim=0cm 0cm 0cm
  0cm]{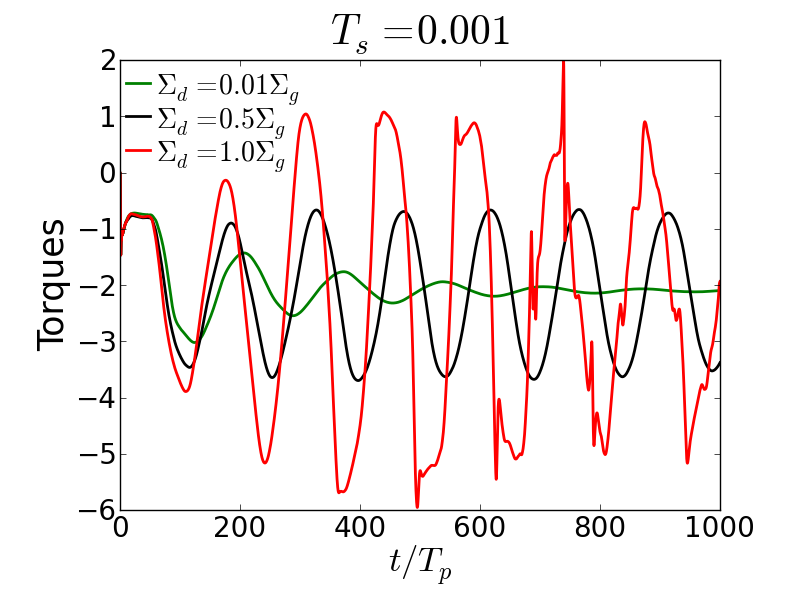}
  \caption{Disc-on-planet torques for a $2M_{\earth}$ planet in a disc
  with stopping time $T_\mathrm{stop}=10^{-3}$ and different initial
  dust-to-gas ratios.  
  \label{torques_vary_d2g}}
\end{figure}

We now examine the effect of dust-loading by varying the metallicity 
$Z$ at fixed  $\Tstop = 10^{-3}$.
In Fig. \ref{torques_vary_d2g}, we compare the decoupled simulation 
with $Z=0.5$ to cases with $Z=0.01$ and $Z=1$. The torque evolution 
for $Z=0.01,\,\Tstop=10^{-3}$ is similar to the fiducial run with $Z=0.5,\,\Tstop=0$. 
This is expected since the baroclinic term, $\left|\nabla\Sigd\times\nabla\Sigg\right|$, vanishes as $Z\to0$.   
From Eq. \ref{dusteq_prim}, we also expect the magnitude of baroclinity to be proportional to
$Z\Tstop$. Indeed, an additional run with $Z=5,\Tstop=10^{-4}$ show similar results to $Z=0.5,\Tstop=10^{-3}$.  



\begin{figure}
\includegraphics[width=\linewidth,clip=true,trim=0cm 1.3cm 0cm
  0cm]{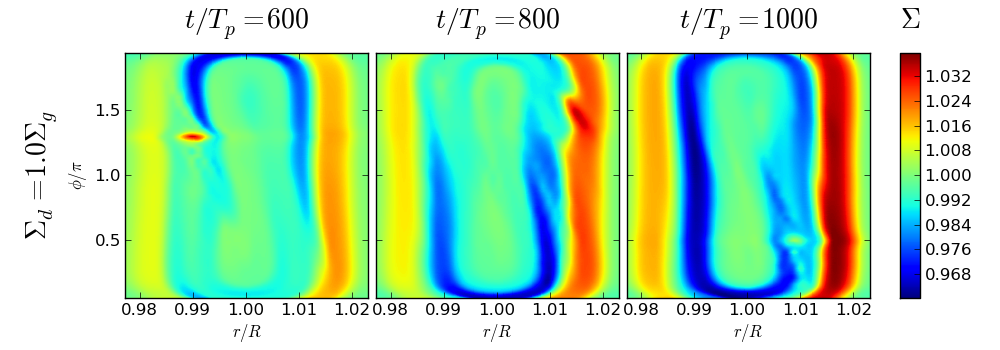}\\
\includegraphics[width=\linewidth,clip=true,trim=0cm 0cm 0cm
  0cm]{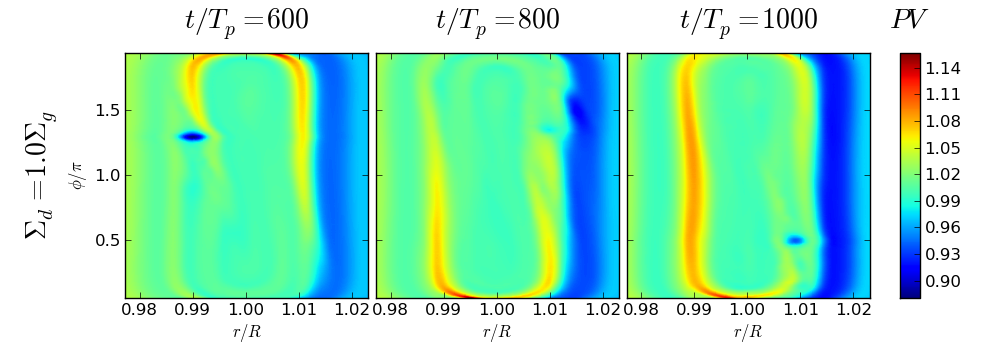}
  \caption{ Disc surface density (top) and potential vorticity
    (bottom) for the high dust-to-gas ratio $Z=1$ run, showing the
    development of a co-orbital vortex (red/blue blob in the surface density/PV, respectively). 
  \label{shot_rho_d2g_high}}
\end{figure}

On the other hand, the torque evolution for the high dust-to-gas ratio 
disc ($Z=1$) attains larger oscillation amplitudes than $Z=0.5$, and
eventually develops high frequency fluctuations on top of the
oscillations at $t\gtrsim 500P_0$. We show in
Fig. \ref{shot_rho_d2g_high} that this is associated with vortex formation
at the separatrix of the co-orbital region. A similar phenomenon was observed by
\cite{paardekooper10b} in their pure-gas, adiabatic simulations. They 
attributed it to the Rossby Wave Instability
\citep[RWI][]{lovelace99,li00} associated with entropy extrema
at the separatrix. This can arise from entropy advection when there is a background entropy gradient. In barotropic discs the  
RWI develops at PV minima \citep[e.g.][see also
  Appendix \ref{dust_trapping}]{valborro07}.

\par
\begin{figure}
\includegraphics[width=\linewidth,clip=true,trim=0cm 0cm 0cm
  0cm]{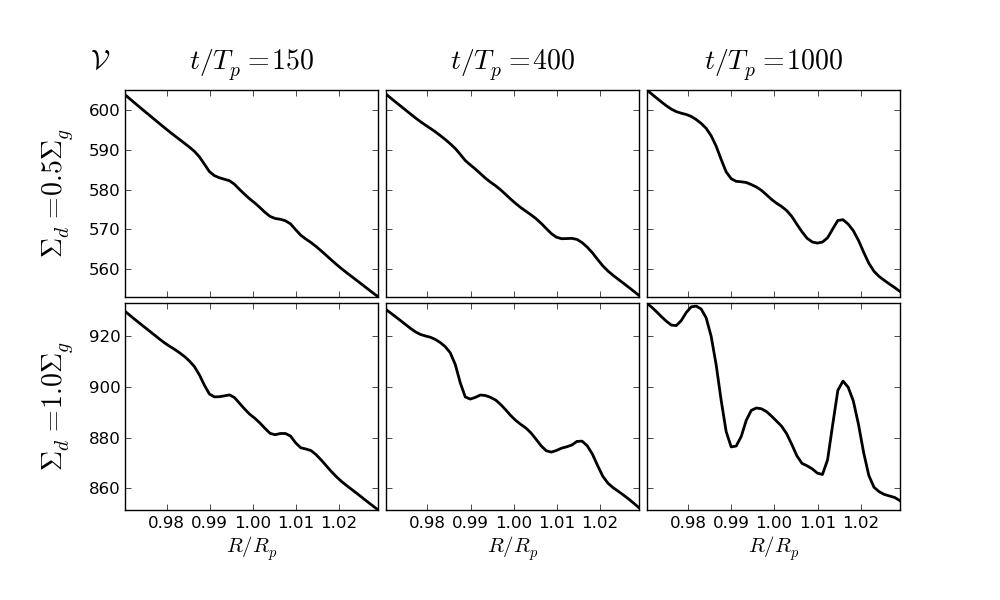}
  \caption{Generalized potential vorticity distribution $\mathcal{V}$ for a $2M_{\earth}$ 
  planet in a dusty disc with particle stopping time 
  $T_\mathrm{stop}=10^{-3}$ and dust-to-gas ratio $Z=0.5$ (upper) and $Z=1.0$ 
  (lower).
  \label{radial_gpv_series_vary_d2g}}
\end{figure}

 
Based on the equivalence between isothermal, dusty gas and adiabatic
pure gas, \citetalias{lin17} argued  the relevant quantity for the 
RWI in isothermal dusty discs is the `generalized potential vorticity'  
\begin{align}
  \mathcal{V} \equiv \frac{\kappa^2}{2\Omega\Sigma}\times\left(1 +
  \frac{\Sigd}{\Sigg}\right)^2, 
\end{align}
where $\kappa^2 = R^{-3}\p_R\left(R^4\Omega^2\right)$ is the square of
         the epicyclic frequency. The generalized PV encapsulates both PV { (the first factor)} and effective entropy { (second term)}. Recall from \S\ref{eff_entropy} that the effective entropy 
of an isothermal, dusty disc is related to the dust-to-gas ratio.  
Fig. \ref{radial_gpv_series_vary_d2g} compares the generalized PV profiles
for $Z=0.5$ and $Z=1$. We indeed find the $Z=1$ case develops deeper
         minima in $\mathcal{V}$,  
which subsequently becomes unstable to vortex formation.  

For our setup the effective entropy is initially uniform because
$\Sigd/\Sigg$ is constant. Extrema in $\mathcal{V}$ at the separatrices thus arise from 
\begin{inparaenum} 
\item advection of the background PV; and/or \label{pv_advect}
\item { baroclinic} PV and effective entropy generation due to
  $\left|\nabla\Sigd\times\nabla\Sigg\right|\neq 0$.   
\end{inparaenum}
To determine which effect is responsible, we ran a case with $Z=1, \Tstop=0$, where baroclinity is absent, 
and found no vortex formation. Thus in the case with $Z=1, \Tstop=10^{-3}$, extrema in $\mathcal{V}$ is
associated with the baroclinic source for PV and effective entropy. 





\subsection{Discs with flat potential vorticity profiles}


Here we consider a disc model with $p=1.5$ so the initial PV
(approximately $\propto R^{3/2 - p}$) is nearly uniform. We choose $q=-0.003$ so the true 
temperature increases slightly outwards to maintain a constant dust-to-gas ratio. 
These parameters imply vanishing co-rotation torques initially.

\begin{figure}
\includegraphics[width=\linewidth,clip=true,trim=0cm 0cm 0cm
  0cm]{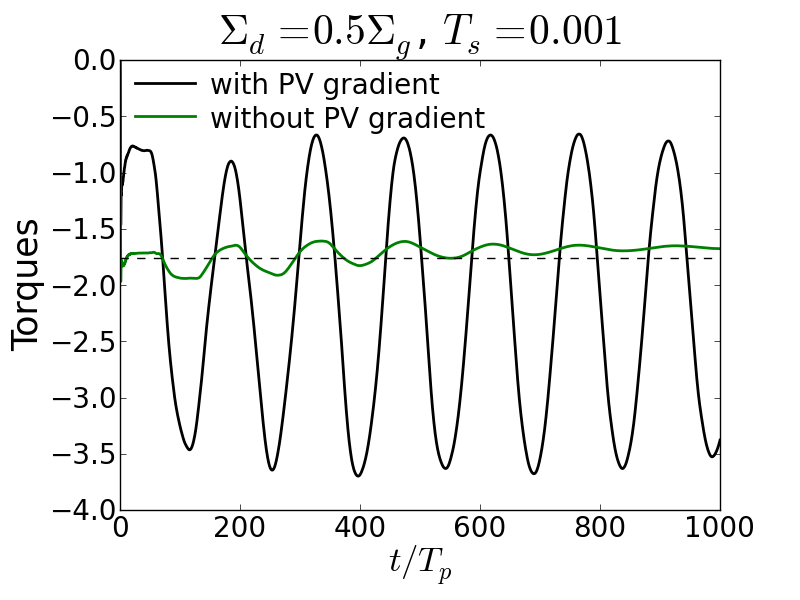}
  \caption{Disc-on-planet torques for a $2M_{\earth}$ planet in a disc
  with stopping time $T_\mathrm{stop}=10^{-3}$ and different initial PV 
  profile. Horizontal lines correspond to semi-analytic Lindblad torques, $\hat{\Gamma}_\mathrm{L}$, for
  locally isothermal discs adapted from \protect\cite{paardekooper10b}.
  \label{torques_flat_pv}}
\end{figure}

In Fig. \ref{torques_flat_pv} we plot the torques measured in this
flat PV run, in comparison with decoupled run with a 
PV gradient, and the analytic Lindblad torque value. The 
oscillation amplitudes are significantly smaller, and the total torque  
remains close to the Lindblad-only values. This experiment
demonstrates that the sustained oscillations requires an 
initial PV gradient, i.e. a non-vanishing initial co-rotation torque
         (specifically the horseshoe drag). 
In that case the initial advection of PV across horseshoe  
turns, thereby creating a PV surplus, feeds the baroclinic source, which further generates PV. 


\subsection{Protoplanetary disc models}\label{ppd}

\begin{figure}
\includegraphics[width=\linewidth,clip=true,trim=0cm 0cm 0cm
  0cm]{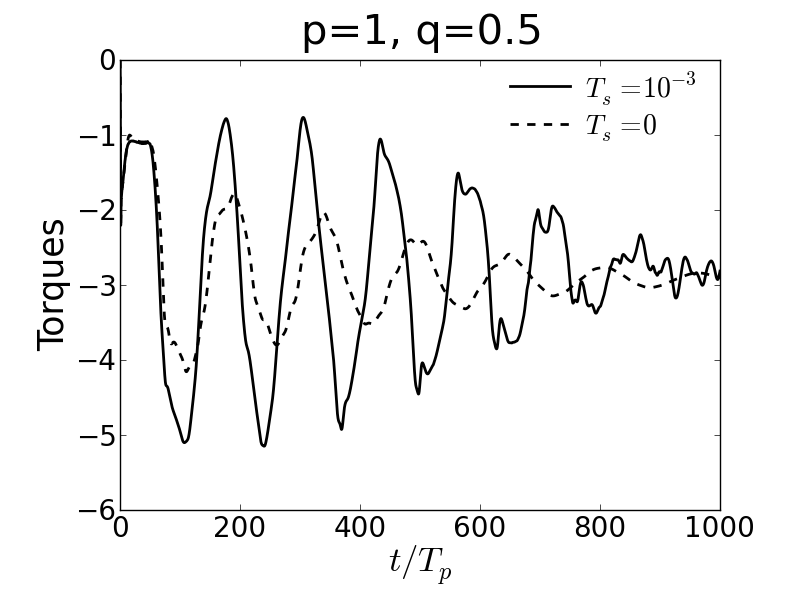}
  \caption{Disc-on-planet torques for a $2M_{\earth}$ planet in a disc
    with $p=1,\,q=0.5$. The initial dust-to-gas ratio and stopping
    time at $R=1$ is $Z=0.5$ and $T_\mathrm{stop}=10^{-3}$ (solid), $T_\mathrm{stop}=0$ (dashed). 
  \label{torques_typical}}
\end{figure}

We now consider a disc model with $p=1$, $q=0.5$, which are typical
         values for protoplanetary discs. In addition to a PV gradient, this model also contains a temperature gradient and
         an initially non-uniform dust-to-gas ratio.
         Fig. \ref{torques_typical} compares the disc-planet torques
         in this model for perfectly coupled and partially coupled
         dust. As before, the $\Tstop=10^{-3}$ run sustains larger amplitude
         torque oscillations than the $\Tstop=0$ run; although 
         oscillations in the former case begins to damp after $t\sim
         400P_0$, unlike the case with $p=q=0$
         (Fig. \ref{torques_extend2}). 

\begin{figure}
\includegraphics[width=\linewidth,clip=true,trim=0cm 1.3cm 0cm
  0cm]{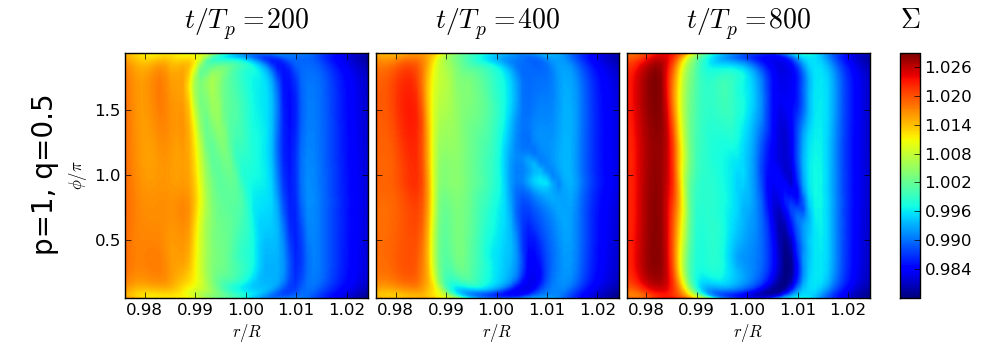}\\
\includegraphics[width=\linewidth,clip=true,trim=0cm 0cm 0cm
  0cm]{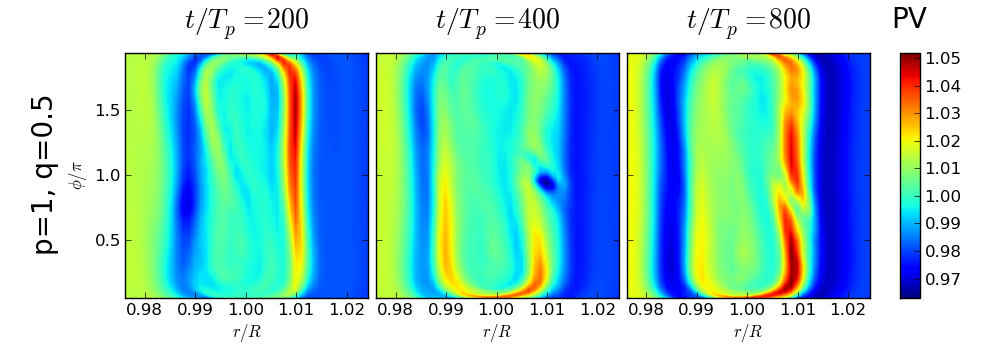}
  \caption{Snapshots of normalized surface density (top) and PV (bottom) for the
    $p=1,\,q=0.5,\,\Tstop=10^{-3}$ 
    case shown in Fig. \protect\ref{torques_typical}. 
  \label{shot_rho_typical}}
\end{figure}

         Interestingly, on longer timescales, Fig.  \ref{shot_rho_typical} show that the PPD
         model with $\Tstop=10^{-3}$ develops a vortex at the
         boundary of the co-orbital region, though this did not occur
         in the $p=q=0$ run.
         The advection of dust-to-gas ratio through 
         horseshoe turns in the presence of a background gradient in
         $\Sigd/\Sigg$ likely contributes to gradients in
         the disc's effective entropy at the co-orbital boundary, in
         addition to entropy generation from baroclinity. This
         effect would favour the RWI in the case with $p=1,\,q=0.5$.

However, unlike the run with high dust-to-gas ratios (\S\ref{dust_load}), here the vortex is a
         transient feature; dissipating as the co-orbital region
         undergoes phase mixing. The total torque then converges to a
         nearly-constant value similar to the $\Tstop=0$ case.



\subsection{Numerical resolution}\label{resolution}


Our simulations are carried out in the inviscid limit. Thus the
         smallest scale is always the grid scale. Numerical 
         diffusion could be important in co-orbital dynamics, specifically the
         separatrices, where large gradients can develop. 

\begin{figure}
\includegraphics[width=\linewidth,clip=true,trim=0cm 0cm 0cm
  0cm]{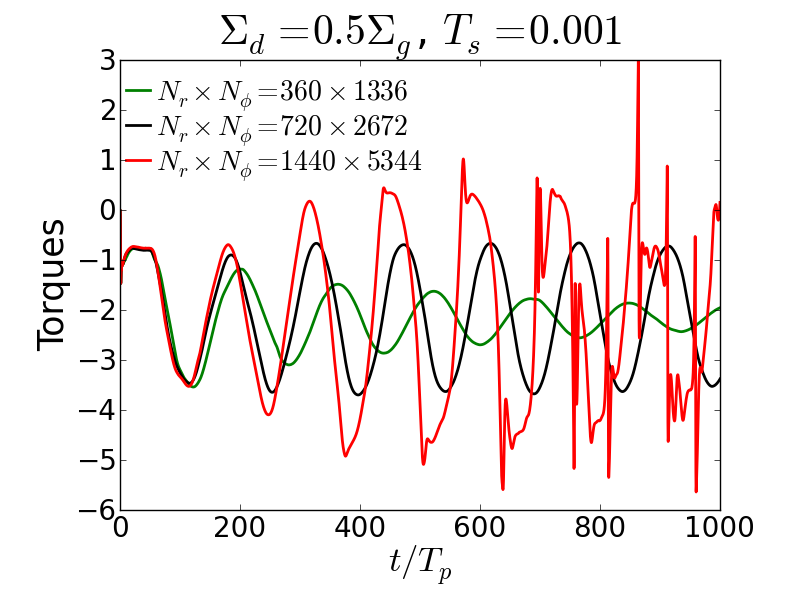}
  \caption{Disc-on-planet torques for a $2M_{\earth}$ planet in a disc
  with stopping time $T_\mathrm{stop}=10^{-3}$ and different
  resolutions.
  \label{torques_vary_res}}
\end{figure}

In Fig. \ref{torques_vary_res} we repeat the decoupled run at half and
double the standard resolution (40, 20 cells per $\widetilde{H}$ in
$r,\phi$, respectively). For $t\lesssim 100P_0$ the linear torques are
well converged. For $t\gtrsim 100P_0$ at low resolution numerical diffusion causes
efficient PV mixing in the co-orbital region, despite baroclinic
source, and the oscillation damps. This effect is mitigated at high 
resolution, resulting in larger torque amplitudes. The torque
magnitudes can exceed the initial values, and even becomes 
positive at times.

However, we also find vortex formation in the high
resolution case due to the RWI, because the large gradients at the
{ separatrices} are better resolved. It is thus unclear if seeking true
         convergence is meaningful in these inviscid simulations. 
We emphasize that the dusty baroclinic generation of PV (and sustained
         co-orbital torques) is a physical effect, but require sufficient resolution to
counter PV mixing/{ numerical diffusion}.  

The different resolutions presented here
         could mimic discs with different levels of a background
         turbulent viscosity. As such, the dust-induced co-orbital torques
         and RWI would only occur in laminar `dead zones' of
         protoplanetary discs. This is consistent with considering a
         high metallicity disc in the first place, since only a low
         level of turbulence would allow dust settling
         \citep{takeuchi02}.


%

\section{Summary and discussion}\label{summary} 

In this paper, we implement 
a new, dust-free model of dusty gas 
developed by \cite{lin17} into the \textsc{pluto}  hydrodynamics 
code. It assumes the dust particles are small enough so that their stopping time
is sufficiently short to treat the dusty gas as a single-phase fluid. The
model allows for a slight decoupling between gas and dust by
accounting for 
dust-gas relative drift via a source term in an effective energy 
equation. 
In this first numerical application of the \citeauthor{lin17}'s
dust-free framework, we utilize the code to revisit disc-planet interaction in the 
limit of high dust-to-gas ratios and focus on disc-planet torques. 
We mainly consider the effect of stopping times $\Tstop$ (or particle size) and
dust-loading $Z$ (or metallicity). 


For perfectly coupled dust ($\Tstop=0$), the disc-planet torques matches to the
torque formula adapted from \cite{paardekooper10b}, which are based on
pure gas simulations. We find a good match when the horseshoe
drag is still unsaturated, as well when the total torque 
converges to the Lindblad-only values ($t\gtrsim1000$ orbits). The effect of dust-loading can be
absorbed into the torque normalisation, namely a reduction in the
effective disc scale-height and in contributing to the total disc
mass. Runs with finite stopping
time but low metallicity (e.g. $\Tstop=10^{-3},\, Z=0.01$) behave
similarly to perfectly coupled dust at higher metallicity{,} in that 
torques immediately damp towards the Lindblad-only values. 


However, for finite stopping time \emph{and} high metallicity,
e.g. $\Tstop=10^{-3}, \, Z = 0.5$, the torque oscillation amplitudes do not damp
until after $\sim 1000$ orbits (cf. the run with perfectly coupled
dust). The torques remains oscillatory until at least $\sim2000$
orbits, although it appears to be converging towards the
Lindblad-values. 
We traced this phenomenon to potential vorticity (PV) generation from the 
misalignment between dust and gas { surface densities} in
the planet's co-orbital region. A PV blob, corresponding to a surface
density `bubble', is created by the initial condition near the
separatrices of the co-orbital region, and experiences a baroclinic PV boost 
whenever it passes by the planet via horseshoe turns. As the PV blob
librates { at} the co-orbital boundary, it introduces periodic 
surface density asymmetries in front of and behind the planet, which
translates to sustained oscillatory co-orbital torques. The PV blob eventually
dissipates, possibly due to numerical diffusion and/or PV mixing in
the co-orbital region. 


Increasing the disc metallicity at fixed stopping time,
e.g. $\Tstop=10^{-3},\, Z=1$, leads to vortex formation in the
co-orbital region. This introduces high frequency variability in the disc
torques. 
In previous, pure gas, adiabatic simulations, co-orbital vortex formation was attributed to the 
Rossby Wave Instability associated with strong entropy gradients at
the separatrices \citep{paardekooper10b}. In general the RWI is
related to extrema in PV and/or entropy profiles. In isothermal, dusty
discs the relevant quantity is the generalised PV, $\mathcal{V} =
(\kappa^2/2\Omega\Sigma)(1 + \Sigd/\Sigg)^2$,
which captures both the PV and effective entropy of isothermal
dusty gas. Analysis of the $\mathcal{V}$ profiles in the unstable
cases show that prior to
vortex formation, extrema in $\mathcal{V}$ indeed develops in the co-orbital
region. 

We also consider two additional disc models: one with zero PV
gradient initially and a protoplanetary disc model with non-zero gradients in
PV, temperature, and initial dust-to-gas ratio. 
In the flat PV disc, the torques immediately jump to the Lindblad torque and
does not change much after. This is consistent with the previous
simulations, which show a PV boost occurs if there is an Eulerian
PV perturbation when a fluid parcel undergoes horseshoe turns. An
initial PV surplus cannot develop if PV is globally flat. This
simulation shows that sustained oscillatory torques require a
background PV gradient. In the protoplanetary disc model, we again
find sustained torque oscillations, but at late times a high frequency
torque component develops due to vortex formation in the co-orbital
region. However, this vortex appears to be transient, dissipating
within the simulation timescale, leaving only the Lindblad torques to
remain. 


\subsection{Implications of dusty disc-planet interaction}

Our main result is that low mass planets in dusty
discs may experience extended periods of oscillatory torques acting
on them. This could lead to growth in the planet's eccentricity. This 
effect requires sufficiently high dust-loading and/or large particles.
If a planet migrates into the inner disc with high dust-to-gas ratio and
large particles --- a typical outcome of particle drift
\citep[e.g.][]{kanagawa17} --- then oscillatory torques may 
develop and introduce planet eccentricity. Alternatively, once
particles in the disc surrounding the planet grows beyond some size, 
finite dust-gas coupling may again introduce oscillatory torques. 


We also find vortex formation in the co-orbital region in some
cases. The effect of co-orbital vortices on vortex formation is less
clear, and is best address through direct numerical
simulations. However, these vortices are expected to act as dust traps
\citep{lyra13}, and may lead to additional planetesimal and hence
planet formation, which would further complicate the orbital migration
of the original planet. 



\subsection{Caveats and outlooks}

Our simulations motivate the study of dust-rich disc-planet
interaction in detail. In order to follow up the possible implications discussed
above, several { improvements needs to be made in future work.}

We constructed special disc profiles in order to initialise our
simulations in exact equilibrium. In fact, we mostly considered discs
without an initial pressure gradient, i.e. no particle drift. In general, however, the disc may be
evolving due to dust-gas interaction 
\citep{kanagawa17}. Future studies should consider more general,
possibly non-equilibrium initial conditions. In particular, 
a global particle and/or gas mass flux across 
the planet's co-orbital region will likely contribute to the
desaturation of co-orbital torques
\citep{mcnally17}. 

In this first study, we fixed the planet's orbit and simply measured the disc-planet
torques. To see how a planet actually migrates in response to these
torques, specifically whether or not it can experience eccentricity
growth, we require a live planet, i.e. an orbit integrator. This
problem, however, will likely require high numerical
resolutions. 


Future studies should also consider fully 3D, stratified disc
models. This is because dust-gas drag --- a key factor in the torque
evolution --- depends on the
local density (Eq. \ref{drag_law}){. In a 3D disc the stopping time increases
  away from the midplane; while the dust-to-gas ratio or local metallicity drops with height.}  
This height dependence may not be properly captured
in our 2D disc models, in which we simply replace the { densities by 
surface densities}. Three-dimensional disc models also allow for explicit
modeling of hydrodynamic turbulence such as that due to the Vertical
Shear Instability \citep{nelson13}. This is expected to be important
for the co-orbital dynamics considered here { as it provides an effective viscosity.}  
However, such 3D 
numerical simulations are prohibitively expensive at the
present time. 


\section*{Acknowledgments}
This work is supported by the Theoretical Institute for Advanced
Research in Astrophysics (TIARA) based in the Academia Sinica
Institute for Astronomy and Astrophysics (ASIAA). 
JWC is supported by the 2017 ASIAA Summer Student Program, where this
project began. All simulations were carried
out on the TIARA High Performance Computing cluster. 

\bibliographystyle{mnras}
\bibliography{ref}

\appendix
\section{Dust-trapping at planetary gap edges}\label{dust_trapping}

We simulate the response of a 2D, razor-thin dusty disc to {  massive
planets such that the disc structure in dust is
significantly perturbed}.  
Here, for the disc parameters we use $p=1$, $q=-0.002$ to obtain a uniform dust-to-gas ratio of 
$\Sigd/\Sigg=0.01$. As before we use $\xi=1.001$, so gas disc is effectively 
isothermal in both structure and thermal response. The reference gas
temperature is such that $h_{\mathrm{g}0}=0.05$. Since $\Sigd\ll\Sigg$, there is no difference between
the true and dynamic (dust-modified) temperature. 
We set the reference stopping
time to $\Tstop=0.007$. This setup is similar to recent explicit dust-plus-gas
simulations performed by \cite{dong17}. However, unlike 
\citeauthor{dong17}, we neglect gas viscosity and
turbulent dust diffusion. 

The disc domain  is $R\in[0.4,2.5]R_0$ and $\phi\in[0,2\pi]$ with
resolution of $N_R\times N_\phi = 384\times 768$. As in
\S\ref{dust_disc_planet}, we use unperturbed boundaries with radial
buffer zones in $R\in[0.4,0.52]R_0$ and $R\in[2.2,2.5]R_0$.  

\subsection{Dust rings}
\begin{figure}
  \includegraphics[scale=.615,clip=true,trim=0cm
  1cm 0cm 1cm]{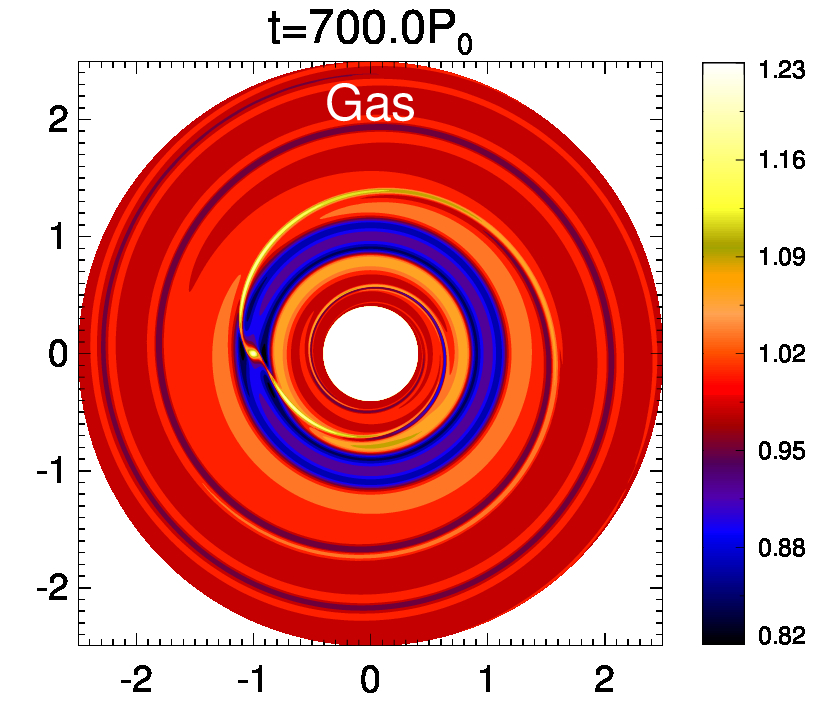}\\
  \includegraphics[scale=.615,clip=true,trim=0cm
  0cm 0cm 1cm]{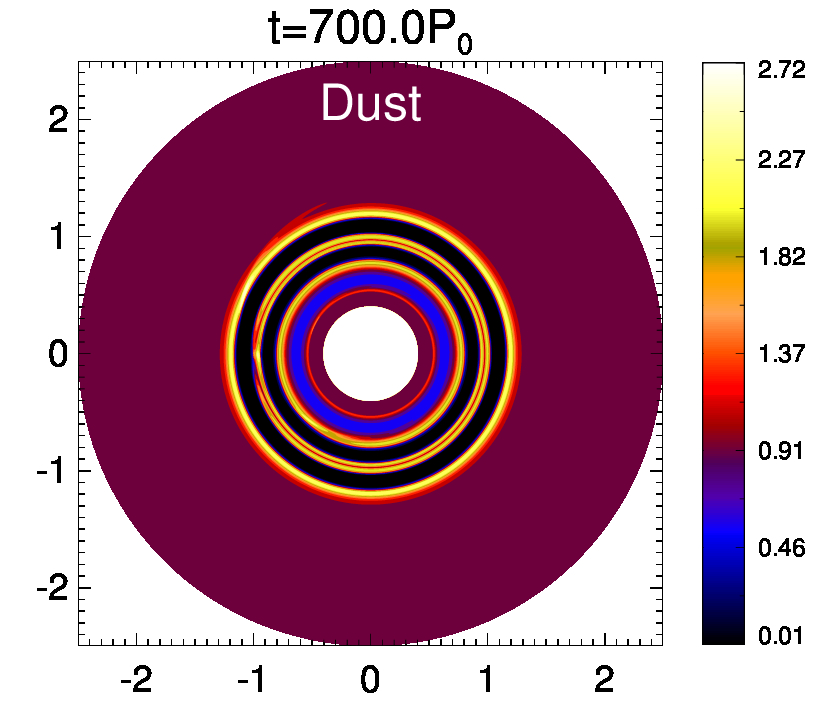}
  \caption{Gas (top) and dust (bottom) response to an embedded planet
    with mass of $10M_{\earth}$ around a star of mass $M_{\sun}$. The
    gas and dust surface densities $\Sigma_\mathrm{d,g}$ are normalized by
    their initial values.\label{dong_dust_ring}
    }
\end{figure}

Fig. \ref{dong_dust_ring} shows the normalised gas and dust surface densities  
at $t=700P_0$ for a planet of mass $M_p=3\times10^{-5}M_*$, or
$M_p=10M_{\earth}$ if $M_*=M_{\sun}$.
While a single, shallow gas gap is
carved with only $\sim 20\%$ depletion in gas;
a double, deep dust gap is formed with almost complete depletion. The dust displays a
multi-ringed structure associated with the gas gap edges and the
co-orbital (horseshoe) region of the planet. In particular, the co-orbital material itself shows a double-ringed structure. These features are very similar to that obtained by \citet[][see their Fig. 1]{dong17} in their two-phase simulations.    

\subsection{Dusty vortex}
\begin{figure}
  \includegraphics[scale=.615,clip=true,trim=0cm
  1cm 0cm 1cm]{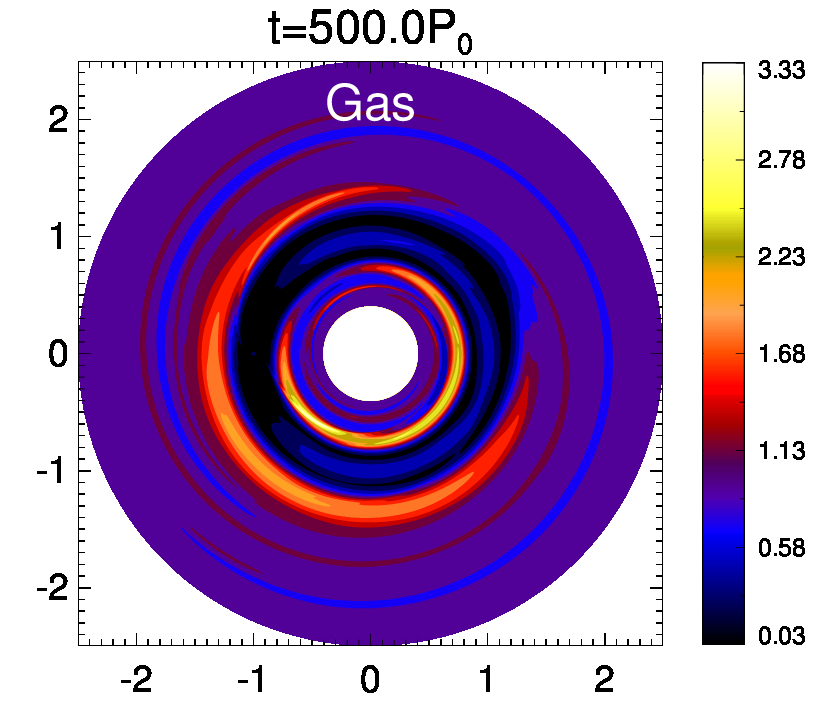}\\
  \includegraphics[scale=.615,clip=true,trim=0cm
  0cm 0cm 1cm]{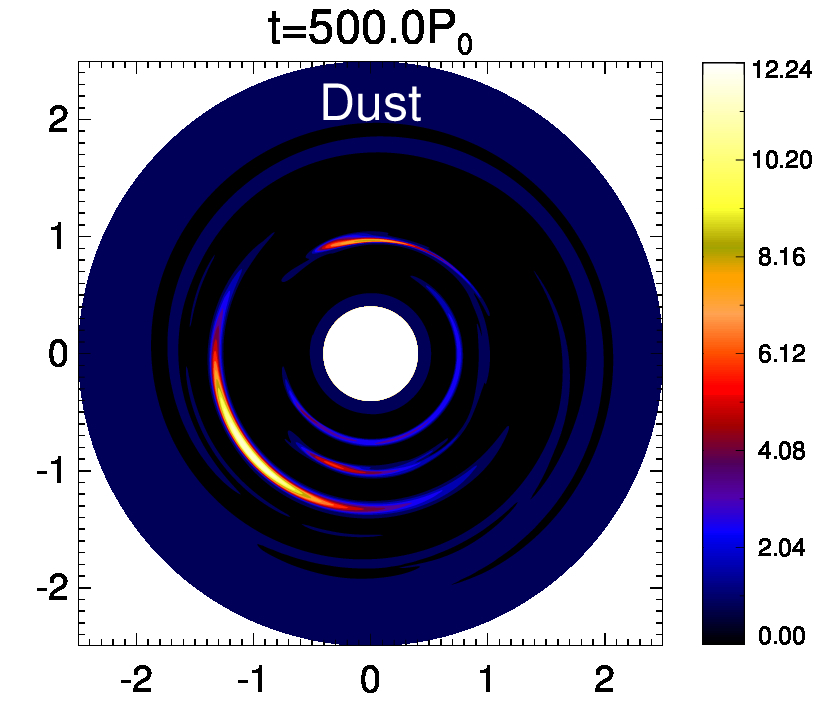}
  \caption{
    Same as Fig. \protect\ref{dong_dust_ring} but with a planet mass
    $M_p=100M_{\earth}$.   
    \label{dong_dust_vortex}
    }
\end{figure}

We repeat the above simulation with $M_p=3\times10^{-4}M_*$; 
equivalent to $100M_{\earth}$ or a Saturn-mass planet around a Solar
mass star. Fig. \ref{dong_dust_vortex} shows the gas and dust 
distribution at $t=500P_0$. For this planet mass a partial gas gap forms, which subsequently 
becomes unstable to the Rossby Wave Instability 
\citep{lin10}, which leads to large-scale vortex formation at the gap
edges. The dust is efficiently trapped by the vortices in both radius and azimuth \citep{lyra13}, 
resulting in a much higher concentration than the gas over-density. The 
dust gap is also wider than the gas gap, probably due to
dust-collection into the outer vortex.  

\section{Streaming instability with one-fluid}\label{streaming}
The streaming instability (SI) is an axisymmetric, linear dust-drag
instability \citep{youdin05a,jacquet11}. SI relies on the
back-reaction of dust-drag on gas, and thus is most efficient in
dust-rich discs ($\rhod\gtrsim\rhog$). This is especially true for
tightly-coupled, small particles with $\Tstop\ll1$ \citep{carrera15,yang16b}.  

Linear SI growth rates may be calculated analytically from a full 
two-fluid treatment of  dusty gas \citep{youdin05a}. In the strong-drag limit,
SI growth rates may also be obtained from the
one-fluid, approximate framework considered here \citepalias{lin17}. This provides a means to test the implementation of 
the diffusion-like term to model dust-gas drag.    


\subsection{Unstratified global equilibria}\label{unstrat_eqm}
In order to compare with the analytical calculations of SI, we
consider axisymmetric, 3D discs but neglect vertical
stratification in the initial equilibrium. This is done by  
ignoring the $z$-dependence of the gravitational
potential. 

The equilibrium density and velocity profiles physically
represent conditions near the disc midplane. We thus set 
\begin{align} 
  \rhog = \frac{\Sigma_\mathrm{g}(R)}{\sqrt{2\pi}\Hgas(R)}\propto R^{-p + q/2 - 3/2},
\end{align}
for consistency with stratified discs considered in Appendix \S\ref{settling}, and recall 
$\Sigma_\mathrm{g}$ is given by Eq. \ref{surfden}. 

The total density is $\rho = \rhog(1 + \epsilon)$, and the
initial dust-to-gas ratio $\epsilon(R)$ is chosen
 to satisfy thermal equilibrium, $ \p_R\left(R \fdust\tstop\p_R
   P\right)=0$,  similarly to the razor-thin disc in
 \S\ref{eqm_struct}. This results in the constraint 
\begin{align}
\frac{\epsilon}{\left(1+\epsilon\right)^2}c_s(R)\rho_\mathrm{g}^{\xi-1}(R)=\mathrm{constant}    
\end{align} 
for power-law discs; assuming $\p_RP\neq0,$ as required for SI (see below).  
We thus set the dust-to-gas ratio via 
\begin{align*}
\frac{\epsilon(R)}{\left[1+\epsilon(R)\right]^2} &= 
\frac{\epsilon_0}{\left(1+\epsilon_0\right)^2} 
\left[\frac{c_{s0}}{c_s(R)}\right]\left[\frac{\rho_\mathrm{g0}}{\rho_\mathrm{g}(R)}\right]^{\xi-1}. 
\end{align*}
Then 
\begin{align}
\frac{d\ln{\left(1+\epsilon\right)}}{d\ln{R}} = \frac{\epsilon}{1 -
  \epsilon}\left\{\frac{q}{2}  
+ (\xi-1)\left[p + \left(\frac{3}{2} - \frac{q}{2}\right)\right]\right\}, 
\end{align} 
The special case of a uniform dust-to-gas ratio is only possible if  
\begin{align}
  q = \frac{2(\xi-1)}{\xi-2}\left(p + \frac{3}{2}\right) \quad
  \text{for constant $\frac{\rhod}{\rhog}$}.\label{pq_relation}
\end{align} 
We choose $\xi=1.001$ and $p=1$, and set $q\simeq -0.005$ in
accordance with Eq. \ref{pq_relation} to obtain a constant $\epsilon$. 

The disc rotation profile is given via 
\begin{align}
  R\Omega^2 &= R\OmK^2 + \frac{1}{\rho}\frac{d P}{d
    R}.
\end{align}
SI requires particle drift in the initial equilibrium, which
translates to a non-zero radial pressure gradient. In the literature it is common use the 
parameterization  
\begin{align} 
  \eta \equiv -\frac{1}{2\rhog R\OmK^2}\frac{\p P}{\p R}. 
\end{align} 
The reference temperature is chosen such that $h_\mathrm{g0} = 0.1$, 
giving a dimensionless pressure gradient $\eta_0\simeq 0.0125$ for the
fiducial disc model. 


\subsection{Simulation setup} 
We initialize the simulation with axisymmetric Eulerian perturbations in $\bm{Q} =
(\rho, \bm{v}, P)$ in the form 
\begin{align}
\delta \bm{Q} = \real \left[\widetilde{\bm{Q}} \exp{\ii \left(k_xR + k_zz - \sigma
  t\right)}\right]. \label{si_pert_form}
\end{align}
Here, $\widetilde{\bm{Q}}$ and $\sigma$ are the complex amplitudes and
frequency, respectively; while $k_{x,z}$ are real wavenumbers. The
dispersion relation $\sigma(k_x,k_z)$ and eigenvector { are} given in 
Appendix D of \citetalias{lin17}. We superimpose
eigenvectors to obtain standing waves in $z$, as described by 
\cite{youdin07b}. We will refer to the dimensionless wavenumber
$K_{x,z}\equiv k_{x,z}\eta_0R_0$ and growth rate $S \equiv
\imag\sigma/\Omega_\mathrm{K0}$.    

We carry out axisymmetric cylindrical $(R,z)$ simulations. 
The domain is $[R_0\pm 2\lambda_x,\pm \lambda_z/2]$, where
$\lambda_{x,z} \equiv 2\pi/k_{x,z}$ is the SI wavelength. We apply
periodic boundaries in $z$, but hold radial ghost zones at their
initial values. In addition, we damp variables toward their initial
values in the innermost and outermost $20\%$ of the radial domain 
on a timescale of $t_\mathrm{damp} = 10^{-2}/s$, where $s$ is the
analytic SI growth rate. This treatment in radius is needed because we   
adopt radially global disc models, while analytic SI solutions 
assume periodicity in $R$. The extended radial domain with damping
minimises the influence of boundary conditions. 


Since SI requires high accuracy, for this problem 
we configure \textsc{Pluto} with piecewise parabolic
reconstruction and third order Runge-Kutta time integration. We use
at least $64$ cells per $\lambda_{x,z}$, corresponding to $ >10^3$
cells per $\Hgas$. 
We impose a minimum dust-fraction $\fdust> 10^{-6}$. 

\subsection{Results}
The simulations are summarized in Table \ref{streaming_runs}. 
These are defined by the initial dust-to-gas ratio $\epsilon_0$,
stopping time $\Tstop$ or $\tau_\mathrm{stop}$, and the 
perturbation wavenumbers $K_{x,z}$. We also give 
analytic growth rates obtained from the one-fluid
framework. We checked that these growth rates are nearly identical to
that obtained from a full two-fluid analysis.  
We measure growth rates in the simulations by 
monitoring maximum deviations from equilibrium in
$\left|R-R_0\right|<\lambda_x$ for 
each fluid variable. We report the mean value $S_\mathrm{sim}$, and 
its spread across the variables, $\Delta S_\mathrm{sim}$. We find good
agreement between simulations and the analytic growth rates.  

\begin{table*}
  \caption{Parameters for the streaming instability. Growth rates $S$ are normalized by
    $\Omega_\mathrm{K0}$. The spread in measured growth rates across variables,
    $\Delta S_\mathrm{sim}$, is normalized by $S_\mathrm{sim}$.  
    The particle stopping time $\tau_\mathrm{stop} =
    T_\mathrm{stop}(1+\epsilon_0).$}  
  \label{streaming_runs} 
  \begin{tabular}{lrlrrrrrr}
    \hline
    Run & $\rhod/\rhog$ & $T_\mathrm{stop}$ ($\tau_\mathrm{s}$) & $K_x, 
    K_z$  & Cells/$\lambda$ & $S$ & $S_\mathrm{sim}$ &  $\Delta S_\mathrm{sim}$  & Comment \\ 
    \hline  
    \hline  
    A  & 3 & 0.025 (0.1) & 30,30 &$64$ &0.4251 & $0.4437$ & $0.5\%$ & 
    Fig. \ref{streaming_runA}---\ref{streaming_runA_2D}
    \\
    \hline
    B & 2 & 0.0033 (0.01) & 80,20 &$128$ &0.1203 & $0.1212$ & $7\%$ &
    Fig. \ref{streaming_runBC_2D}
    \\
    \hline
    C & 3 & 0.001 (0.004) & 160,40 &$128$ &0.1059 & $0.1134$ &
    $12\%$ & Fig. \ref{streaming_runBC_2D} \\  
    \hline
  \end{tabular}
\end{table*}

\begin{figure}
  \centering
  \includegraphics[width=\linewidth]{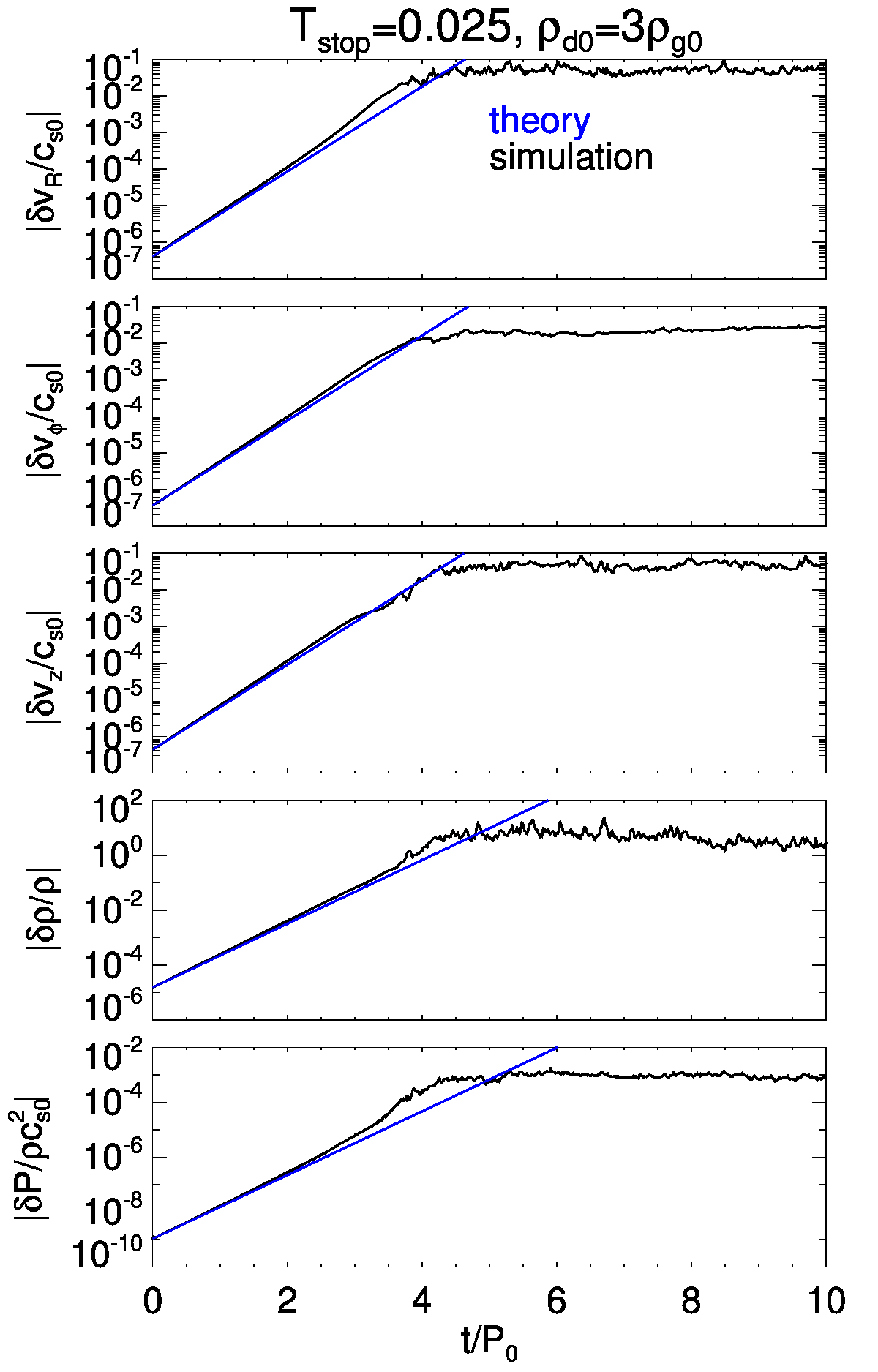}
  \caption{Growth of the streaming instability in run A (see Table
    \ref{streaming_runs}). The evolution in the maximum deviation from
    initial conditions are plotted (black) and compared to the
    expected linear growth of SI (blue). 
    \label{streaming_runA}
    }
\end{figure}

\begin{figure}
  \centering
  \includegraphics[width=\linewidth,clip=true,trim=0cm 1.8cm 0cm
  .9cm]{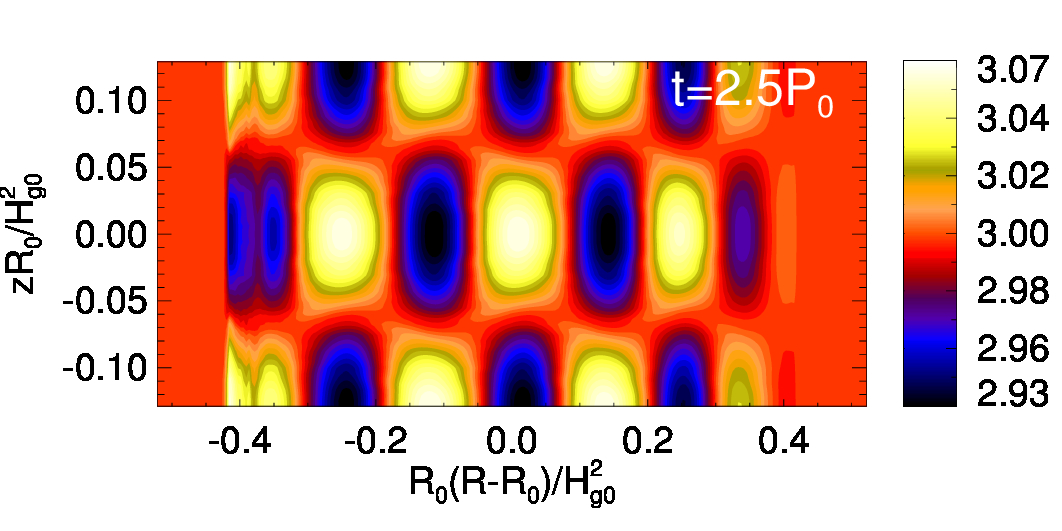}\
  \includegraphics[width=\linewidth,clip=true,trim=0cm 1.8cm 0cm
  .9cm]{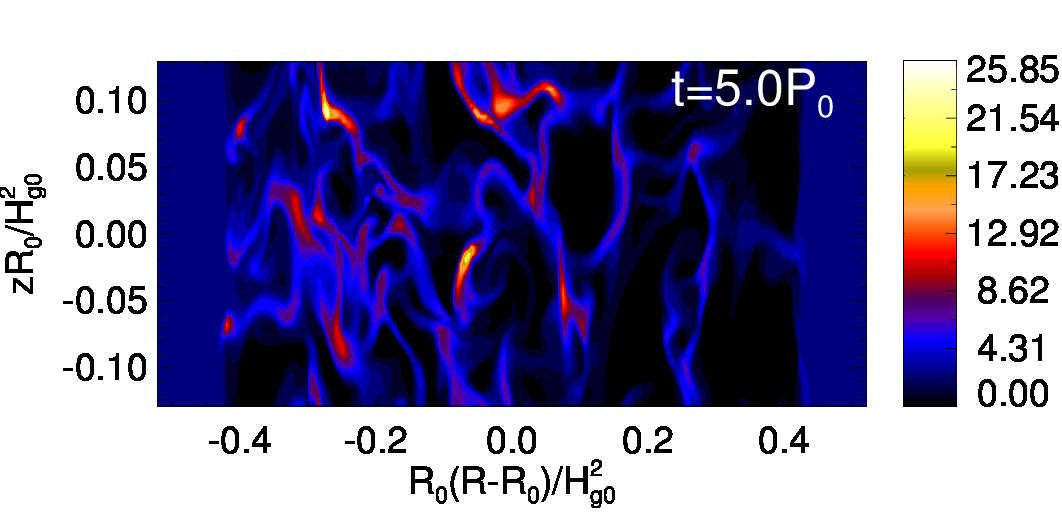}\\
  \includegraphics[width=\linewidth,clip=true,trim=0cm 0cm 0cm
  .9cm]{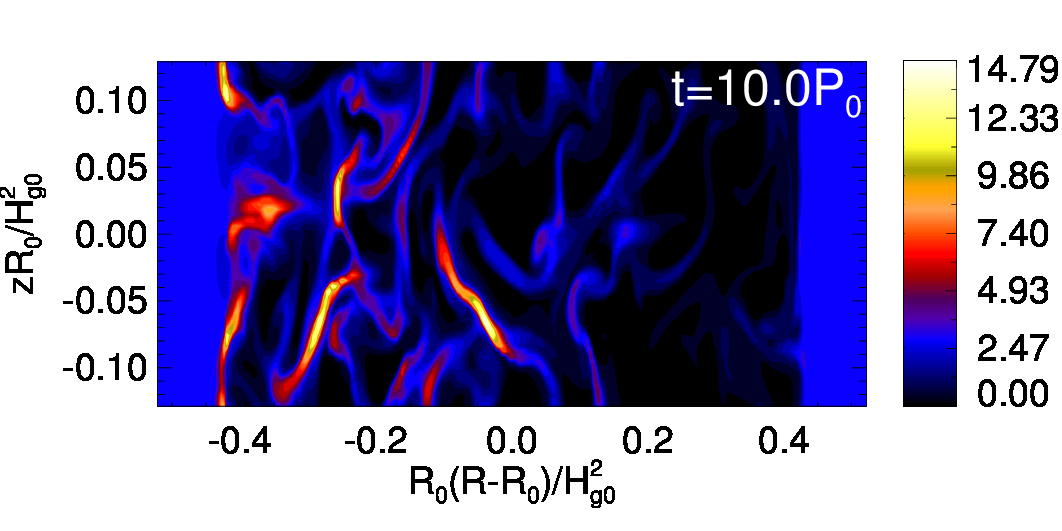}  
  \caption{Evolution of dust-to-gas ratio for run
    A with $T_\mathrm{stop}=0.025$ ($\tau_\mathrm{s}=0.1$) at $2.5$
    orbits (top, end of linear phase), $5$ orbits 
    (middle, saturated state) and $10$ orbits (bottom, end of
    simulation). Notice the inwards 
    migration of dust clumps due to the global radial pressure gradient.  
    \label{streaming_runA_2D}
    }
\end{figure}

Run A is same as the `linA' code test used by other authors to
benchmark their dusty-gas  algorithms
\citep{youdin07b,balsara09,miniati10,bai10b,yang16}. 
Fig. \ref{streaming_runA} show exponential growth of hydrodynamic 
variables and Fig. \ref{streaming_runA_2D} show 
evolution of the dust-to-gas ratio. In the saturated state the 
dust-to-gas ratio ranges from zero to few 10's, showing strong
particle clumping. This is consistent with previous numerical
simulations with similar dust parameters \citep[][their run
`AC']{johansen07}.

\begin{figure}
  \centering
  \includegraphics[width=\linewidth,clip=true,trim=0cm 1.3cm 0cm
    0cm]{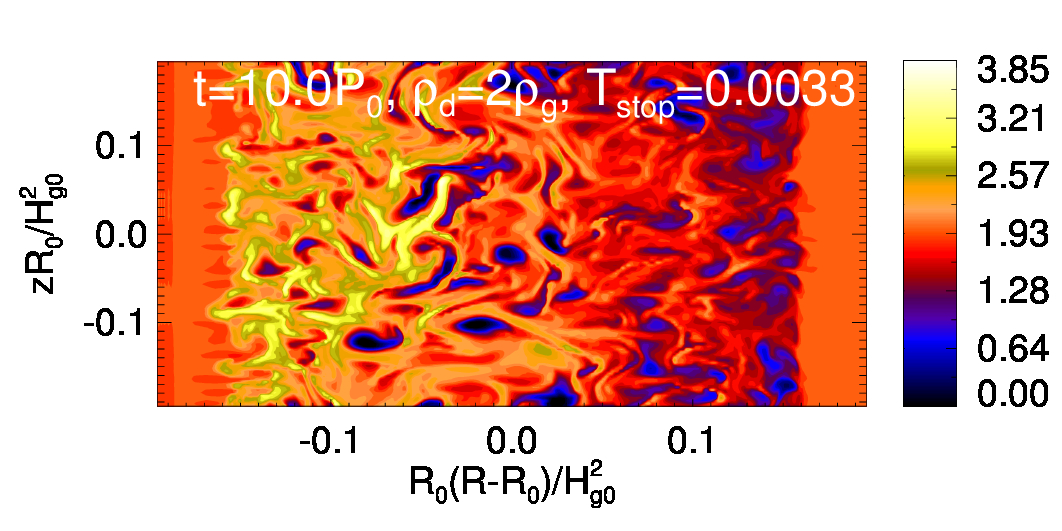} 
  \includegraphics[width=\linewidth,clip=true,trim=0cm 0cm 0cm 0.5cm]{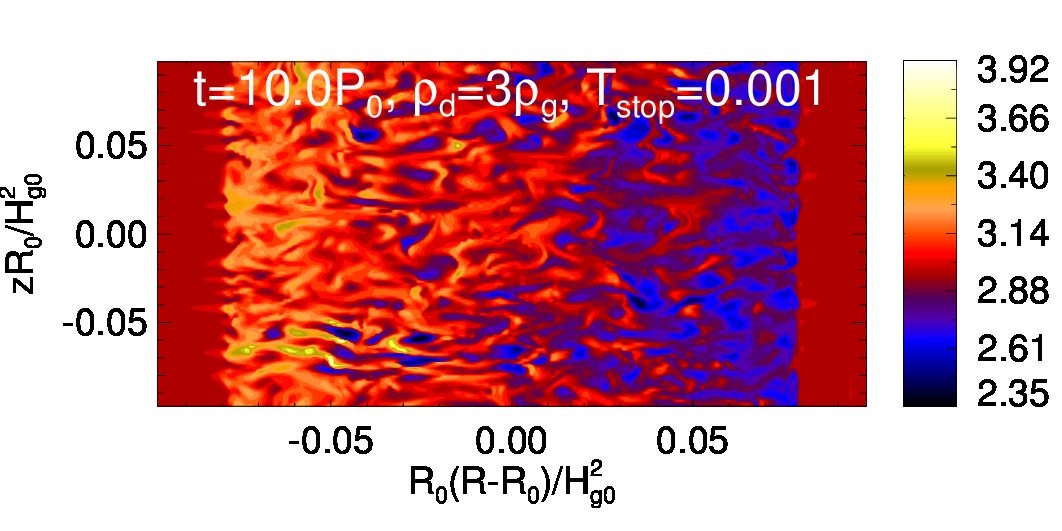}
  \caption{Dust-to-gas ratio at the saturated state of SI in run
    B with $\rhod/\rhog = 2, T_\mathrm{stop} = 0.0033$ (top) and C
    with  $\rhod/\rhog = 3, T_\mathrm{stop} = 0.001$ (bottom).
    \label{streaming_runBC_2D}
    }
\end{figure}

Runs B and C adopt smaller $\Tstop$ and are shown in
Fig. \ref{streaming_runBC_2D}. 
We again find agreement with analytic growth rates  
in the linear phase, although the spread in $S_\mathrm{sim}$ increases
with smaller $\Tstop$. In the saturated state, Run B ($\Tstop=0.0033$) 
still develop voids with $\rhod/\rhog\sim 0$, but Run C ($\Tstop=0.001$) only
develop { order-unity} fluctuations in $\rhod/\rhog$. Neither runs show strong
particle clumping within the simulation timescale. This is consistent
with recent simulations of SI for small particles carried out by
\cite{yang16b}. These authors find small particles only clump after
integration timescales $\gtrsim O(10^2)$ orbits.  


\section{Dust settling in protoplanetary discs}\label{settling}
We now consider vertically stratified discs.
In the presence of vertical stellar gravity, dust settles towards   
the midplane of a protoplanetary disc on a timescale of $t_\mathrm{settle}\sim 1/\Omega^2\tstop$
\citep{takeuchi02}. For $\Tstop=10^{-3}$ we expect $t_\mathrm{settle}\simeq 160$ orbits.  
We demonstrate our hydrodynamic model of dusty gas reproduces this basic result.   

\subsection{Stratified disc model}
We initialize the disc with an axisymmetric, steady state defined by 
\begin{align}
  R\Omega^2 &= R\OmK^2\left(1 - \frac{3}{2}\frac{z^2}{R^2}\right) + \frac{1}{\rho}\frac{\p P}{\p
    R},\label{rad_eqm3D}\\ 
  0 & = z\OmK^2 + \frac{1}{\rho}\frac{\p P}{\p z}. \label{vert_eqm3D}
\end{align}
Given a dust-to-gas ratio distribution $\epsilon(R,z)$, we may solve
Eq. \ref{rad_eqm3D}---\ref{vert_eqm3D} to obtain $\rho(R,z)$ and
$\Omega(R,z)$. We use the thin-disk approximation 
for the
star potential (Eq. \ref{thin_pot}) in order to obtain analytic 
expressions below.  

We { set} the initial dust-to-gas ratio { to} 
\begin{align}
 \epsilon(R,z) = \epsilon_\mathrm{mid}(R) \times \exp{\left( -
  \frac{z^2}{2H_\epsilon^2}\right)}.\label{dg3D}
\end{align}  
Here and below the subscript `mid' indicates the midplane value, taken
from the unstratified profiles described in Appendix \ref{unstrat_eqm}. 
The characteristic thickness in the dust-to-gas ratio, $H_\epsilon$,
is defined via  
\begin{align}
 \frac{1}{H_\epsilon^2} \equiv \frac{1}{\Hdust^2} -  \frac{1}{\Hgas^2}, 
\end{align}
where the dust layer thickness $\Hdust$ is parameterized by $\delta
\equiv \Hdust/\Hgas < 1$. As $\delta\to 1$ the dust becomes perfectly
mixed with the gas, giving a vertically uniform dust-to-gas ratio.   

Using Eq. \ref{dg3D} we can integrate Eq. \ref{vert_eqm3D} (taking { the polytropic index} $\xi=1$) in $z$ to  
obtain    
\begin{align}
  \rhog  = \rho_\mathrm{g,mid}\exp{\left\{-\frac{z^2}{2\Hgas^2} -
  \epsilon_\mathrm{mid}\frac{H_\epsilon^2}{\Hgas^2}\left[ 1-
  \exp{\left(-\frac{z^2}{2H_\epsilon^2}\right)}\right]\right\}}.
\end{align}
The gas vertical structure is effectively Gaussian, since the
$z$-dependence in the second term is weak (especially for well-mixed
dust with $H_\epsilon\to\infty$). The total density is
$\rho~=~\rhog(1+~\epsilon)$ with $\epsilon$ given by Eq. \ref{dg3D}.   

The disc orbital frequency is obtained from Eq. \ref{rad_eqm3D},
\begin{align}
&\Omega(R,z) = \OmK\left[1  - \frac{3}{2}\frac{z^2}{R^2} + 
    \frac{P}{\rho\left(R\OmK\right)^2}\left(\xi
  \frac{\p\ln{\rho_\mathrm{g}}}{\p\ln{R}} - q\right)\right]^{1/2}.
\end{align}
Note that there is generally vertical shear, $\p_z\Omega\neq0$. This arises
from a radial temperature gradient and/or a radial gradient 
in the dust-to-gas ratio \citepalias{lin17}. 


By construction these initial conditions do not satisfy thermal
equilibrium, 
\begin{align*}
  \frac{1}{R}\frac{\p}{\p R}\left(R \fdust\tstop \frac{\p P}{\p
      R}\right) +\frac{\p}{\p z}\left(\fdust\tstop \frac{\p P}{\p 
      z}\right) \neq 0. 
\end{align*}
However, since $\epsilon_\mathrm{mid}$ is chosen to minimize radial
dust evolution (Appendix 
\ref{unstrat_eqm}), this setup allows us to focus on vertical dust
settling.


\subsection{Simulations}
For this problem we adopt an axisymmetric spherical grid $(r,\theta)$
with $r\in[0.6,1.4]R_0$, $\tan{(\pi/2-\theta)}\in[-2,2]h_\mathrm{g0}$,
and $h_\mathrm{g0}=0.05$. We use a resolution of $N_r \times N_\theta
= 840\times 240$, corresponding to  about $50$ to $60$ cells per
$\Hgas$ at unit radius. We use logarithmic spacing in $r$ but uniform
spacing in $\theta$.  { We return to linear reconstruction and 
second order Runge-Kutta time integration, as used in the main text.} 

For the initial disc profile we use $p=1$, $q\simeq -0.005$ and
$\Hdust=0.99\Hgas$ so the initial dust-to-gas ratio $\epsilon \simeq
0.01$ is nearly uniform both radially and vertically.
We choose $|q|\sim 0$ to minimise 
vertical shear and  associated instabilities \citep{nelson13,barker15,lin15}, which 
would stir up particles and oppose dust settling \citep{flock17}. 
We taper $\epsilon$ to zero near radial boundaries in the initial
setup, and apply reflective boundary conditions in both $r$ and
$\theta$.   

\begin{figure}
  \centering
  \includegraphics[width=\linewidth,clip=true,trim=0cm 1.8cm 0cm
  0.95cm]{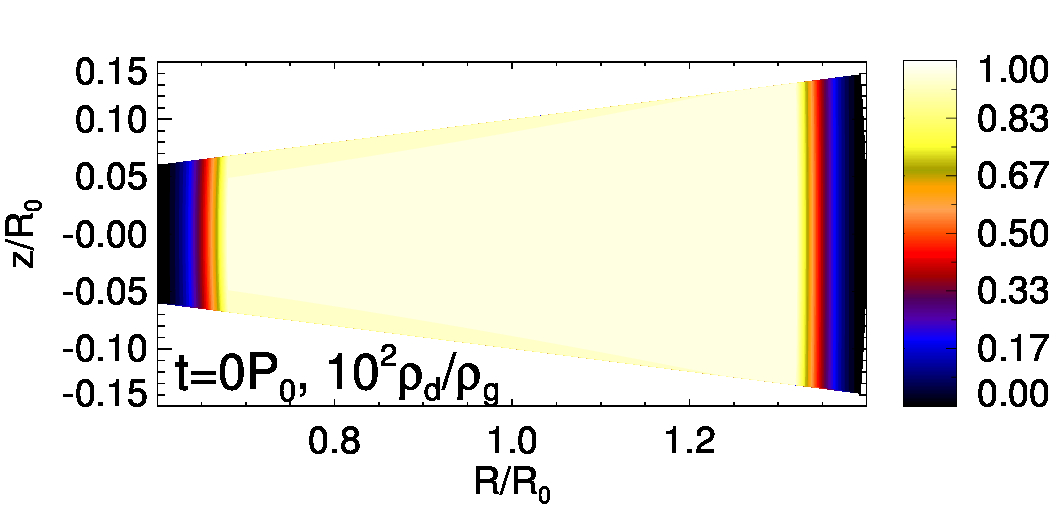}
  \includegraphics[width=\linewidth,clip=true,trim=0cm 1.8cm 0cm
  0.95cm]{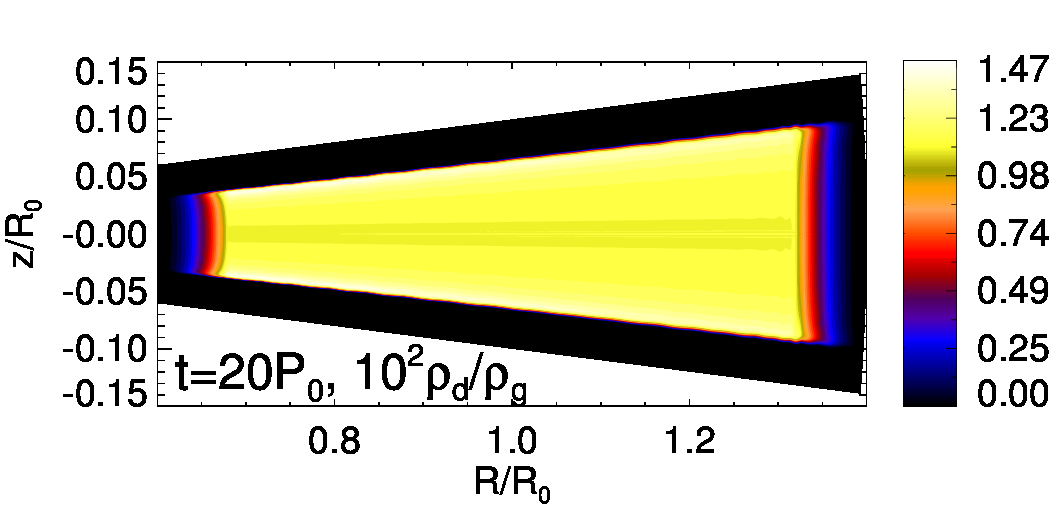}\\
  \includegraphics[width=\linewidth,clip=true,trim=0cm 0cm 0cm 0.95cm]{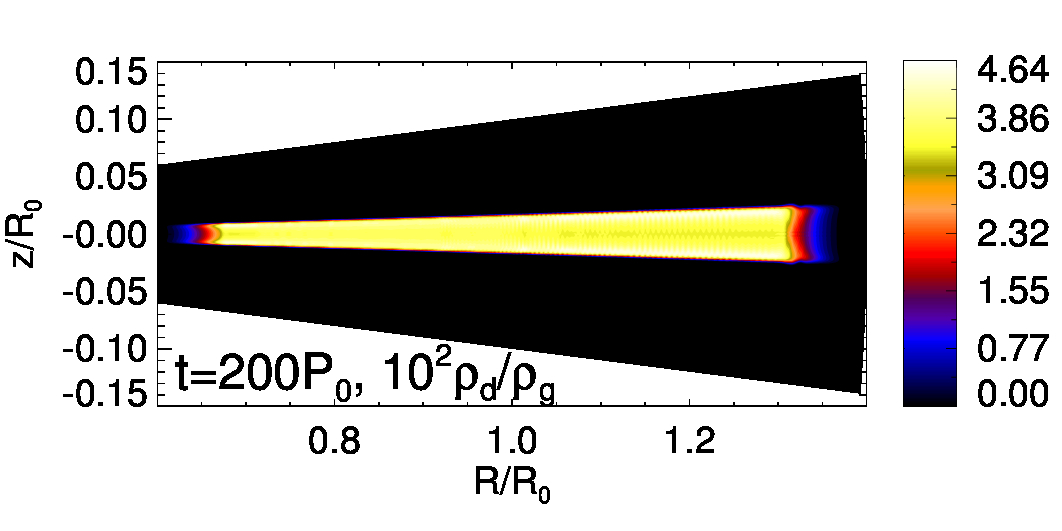}
  \caption{Dust-settling in a vertically stratified disc. The
    dust-to-gas ratio (scaled by $100$) is shown.
	 The reference stopping time is $\Tstop=10^{-3}$. \label{dust_settle}
    }
\end{figure}

Fig. \ref{dust_settle} show the evolution of $\rhog/\rhod$. As 
expected the dust settles to a thin layer ($\sim 0.2\Hgas$) on a
timescale of $\sim 100$ orbits. Settling occurs outwards since 
more dynamical times have elapsed at smaller radii at a given snapshot.
Notice early in 
the settling process the dust-to-gas ratio peaks near the 
surface of the dust layer. These results are similar to dust-settling  
simulations performed by \cite{price15}. 

\end{document}